\documentstyle[12pt,epsfig]{article}

\def\as{\alpha_{\mbox{\tiny S}}}
\def\b0{\beta_0}

\def\bom#1{{\mbox{\boldmath $#1$}}}
\def\ee{e^+e^-}
\def\eps{\epsilon}
\def\gtap{\raisebox{-.4ex}{\rlap{$\sim$}} \raisebox{.4ex}{$>$}}

\def\quart{{\textstyle {1\over4}}}
\def\thrhf{{\textstyle {3\over2}}}
\def\ycut{y_{\mbox{\scriptsize cut}}}

\def\beq{\begin{equation}}
\def\eeq#1{\label{#1}\end{equation}}
\def\eeqn{\end{equation}}

\def\beqa{\begin{eqnarray}}
\def\eeqa#1{\label{#1}\end{eqnarray}}
\def\eeqan{\end{eqnarray}}

\let\bar=\overbar

\def\VEV#1{\left\langle{ #1} \right\rangle}

\def\half{\frac{1}{2}}

\def\Dslash{\not{\hbox{\kern-4pt $D$}}}
\def\dslash{\not{\hbox{\kern-2pt $\del$}}}

\def\ee{e^+e^-}

\def\msb{{\bar{\ssstyle M \kern -1pt S}}}

\def\Title#1{\begin{center} {\Large #1 } \end{center}}

\begin{document}
\begin{flushright}CERN-TH/99-387\\Cavendish-HEP-99/16\end{flushright}

\Title{\bf Fragmentation and Hadronization\footnote{Plenary talk
at XIX International Symposium on Lepton and Photon Interactions
at High Energies, Stanford University, August 1999.}}

\bigskip\bigskip


\begin{raggedright}  

{\it B.R.\ Webber\index{Webber, B.R.}\\
Theory Division, CERN, 1211 Geneva 23, Switzerland, and\\
Cavendish Laboratory, University of Cambridge, Cambridge CB3 0HE,
U.K.\footnote{Permanent address.}}
\bigskip\bigskip
\end{raggedright}

\section{Introduction}\label{sec:intro}
Hadronic jets are amongst the most striking phenomena in
high-energy physics, and their importance is sure to persist
as searching for new physics at hadron colliders becomes the
main activity in our field.  Signatures involving jets almost
always have the largest cross sections but are the most
difficult to interpret and to distinguish from background.
Insight into the properties of jets is therefore doubly
valuable: both as a test of our understanding of strong
interaction dynamics and as a tool for extracting new
physics signals in multi-jet channels.

In the present talk I shall concentrate on jet fragmentation
and hadronization, the topic of jet production having been
covered admirably elsewhere \cite{Womersley,Mangano:1999sz}.
The terms fragmentation and hadronization are often used interchangeably,
but I shall interpret the former strictly as referring
to inclusive hadron spectra, for which factorization
`theorems'\footnote{I put this term in inverted commas because
proofs of factorization in fragmentation do not really extend
beyond perturbation theory \cite{Collins:1987pm}.}
are available. These allow predictions to be made without
any detailed assumptions concerning hadron formation.
A brief review of the relevant theory is given in
sect.~\ref{sec:frag}.

Hadronization, on the other hand, will be taken here to
refer specifically to the mechanism by which quarks and
gluons produced in hard processes form the hadrons that are
observed in the final state.  This is an intrinsically
non-perturbative process, for which we only have models at present.
The main models are reviewed in sect.~\ref{sec:hadro}.
In sect.~\ref{sec:yields} their predictions, together with
other less model-dependent expectations, are compared with
the latest data on single-particle yields and spectra.

One of the most important objectives of jet studies is to understand
the differences between jets initiated by different types of partons,
especially quarks versus gluons, which could also be valuable in
new physics searches. The wealth of recent data on quark-gluon jet
differences is discussed in sect.~\ref{sec:qg}.  Another goal is to compare
jets produced in different processes such as $\ee$ annihilation and
deep inelastic scattering (DIS). This is being done by the H1 and
ZEUS collaborations at HERA; some of their results are discussed in
sect.~\ref{sec:DIS}.

A good understanding of heavy quark jets is especially important
since these are expected to be copiously produced in new processes
such as Higgs boson decay. Section \ref{sec:heavy} discusses some
recent results on heavy quark fragmentation.

Finally, sects.~\ref{sec:BE} and \ref{sec:WW} deal with
correlation effects, first those of Bose-Einstein
origin and then those expected in fully hadronic
final states from $\ee\to W^+W^-$. A summary of the main
points is given in sect.~\ref{sec:conc}.

Regrettably I have no space to discuss many other
relevant and interesting topics, such as:
fragmentation function parametrizations \cite{Jakob};
photon fragmentation function;
polarization in hadronization;
event shapes and power corrections \cite{Beneke};
fluctuations and intermittency;
dynamical two-particle correlations;
heavy quark production in jets ($g\to c\bar c, b\bar b$);
transverse energy flow in deep inelastic scattering;
underlying event in DIS and hadron-hadron collisions;
identified particle production in DIS;
tests of QCD coherence;
jet profiles and substructure \cite{Womersley}. Apologies also to
all those whose work I have omitted or mentioned only superficially. 

In citing the latest experimental data (mostly still preliminary)
I have relied on the reference system of the EPS-HEP99 conference in
Tampere, Finland \cite{Tampere}, since few experimental papers were
explicitly submitted to LP99.  In the case of the large collaborations,
papers submitted to EPS-HEP99 and/or LP99 can usually be found easily
via the collaboration web pages \cite{ALEPH}--\cite{D0}.

\section{Jet fragmentation -- theory}\label{sec:frag}
\begin{figure}\begin{center}
\epsfig{file=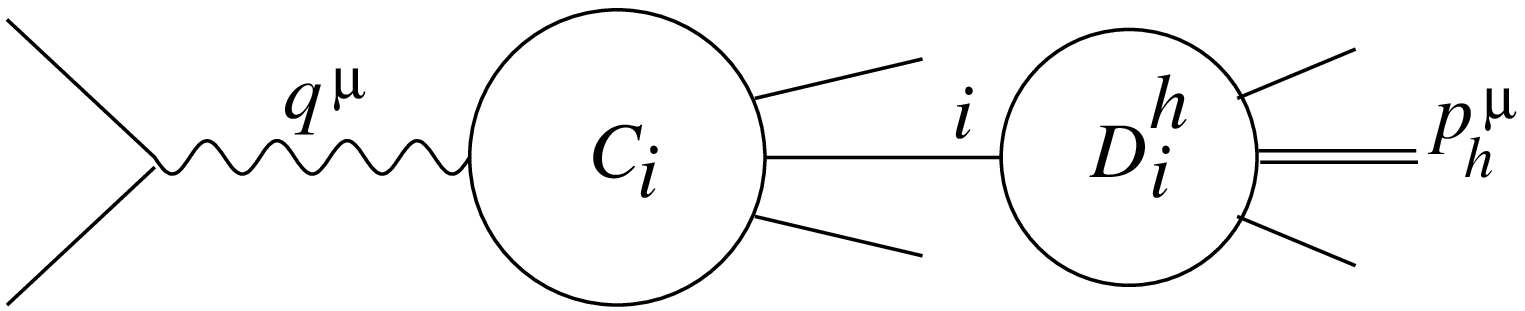,width=8cm}
\caption{Factorization structure of $\ee\to hX$.}
\label{fig:factn}\end{center}\end{figure}

Let us start by recalling the basic factorization structure of the
single-particle inclusive distribution, e.g.\ in $\ee\to hX$
(fig.~\ref{fig:factn}):
$$
F^h(x,s) = \sum_i\int_x^1\frac{dz}{z}C_i(z,\as(s))D^h_i(x/z,s)
$$
$$s=q^2\;,\qquad\qquad x=2p_h\cdot q/q^2 = 2E_h/E_{cm}$$
where $C_i$ are the coefficient functions for this particular process
(including all selection cuts etc.) and $D^h_i$ is the {universal}
{fragmentation function} for parton $i\to$ hadron $h$.

The fragmentation functions are not perturbatively calculable but their
$s$-dependence ({scaling violation}) is given by the DGLAP equation:
$$
s\frac{\partial}{\partial s} D^h_i(x,s)
= \sum_j\int_x^1\frac{dz}{z}P_{ji}(z,\as(s))D^h_j(x/z,s)
$$
Thus they can be parametrized at some fixed scale $s_0$ and then predicted
at other energies \cite{Jakob}.

In certain kinematic regions, higher-order corrections are enhanced by
large logarithms, which need to be resummed.
At small $x$, $\log x$ enhanced terms can be
resummed by changing the DGLAP equation to
$$
s\frac{\partial}{\partial s} D^h_i(x,s)
= \sum_j\int_x^1\frac{dz}{z}P_{ji}(z,\as(s))D^h_j(x/z,z^2s)
$$
This is commonly known as the modified leading-logarithmic approximation
(MLLA) \cite{Azimov:1986by,Fong:1989qy,Dokshitzer:1991wu}.
The effect of resummation is to generate a characteristic hump-backed shape 
in the variable $\xi =\ln(1/x)$, with a peak at $\xi_p\sim\quart\ln s$.

Large logarithms of ratios of invariants may also appear inside
the coefficient functions $C_i$, for example in three-jet events when
the angles between jets become small. In some cases these can be absorbed
into a change of scale in the fragmentation functions.
Examples will be encountered in sects.~\ref{sec:yields} and \ref{sec:qg}. 

Although universal, fragmentation functions are factorization scheme dependent.
The splitting functions $P_{ji}$ are also scheme dependent in higher orders. 
To specify the scheme requires calculation of the coefficient functions to
(at least) next-to-leading order.  This has only been done in a few cases.
Thus there is need for theoretical work to make full use of the data on
fragmentation functions.

\section{Hadronization Models}\label{sec:hadro}
\subsection{General ideas}
\noindent{\em Local parton-hadron duality} \cite{Azimov:1985np}.
 Hadronization is long-distance process, involving only small
 momentum transfers. Hence the flows of energy-momentum and
 flavour quantum numbers at hadron level should follow those at
 parton level. Results on inclusive spectra and multiplicities
 support this hypothesis.

\noindent{\em Universal low-scale $\as$}
\cite{Dokshitzer:1995zt,Dokshitzer:1996ev,Dokshitzer:1996qm}.
 Perturbation theory works well down to low scales, $Q\sim 1$ GeV.
 Assume therefore that
 $\as(Q^2)$ can be defined non-perturbatively for all $Q$, and use it
 in evaluation of Feynman graphs. This approach gives a good description
 of heavy quark spectra and event shapes.

\subsection{Specific models}
The above general ideas do not try to describe the mechanism of
hadron formation.  For this we must so far resort to models.
The main current models are {\em cluster} and {\em string}
hadronization. We describe briefly the versions used in the
HERWIG and JETSET event generators, respectively.
\begin{figure}\begin{center}
\begin{minipage}{50mm}
\epsfig{file=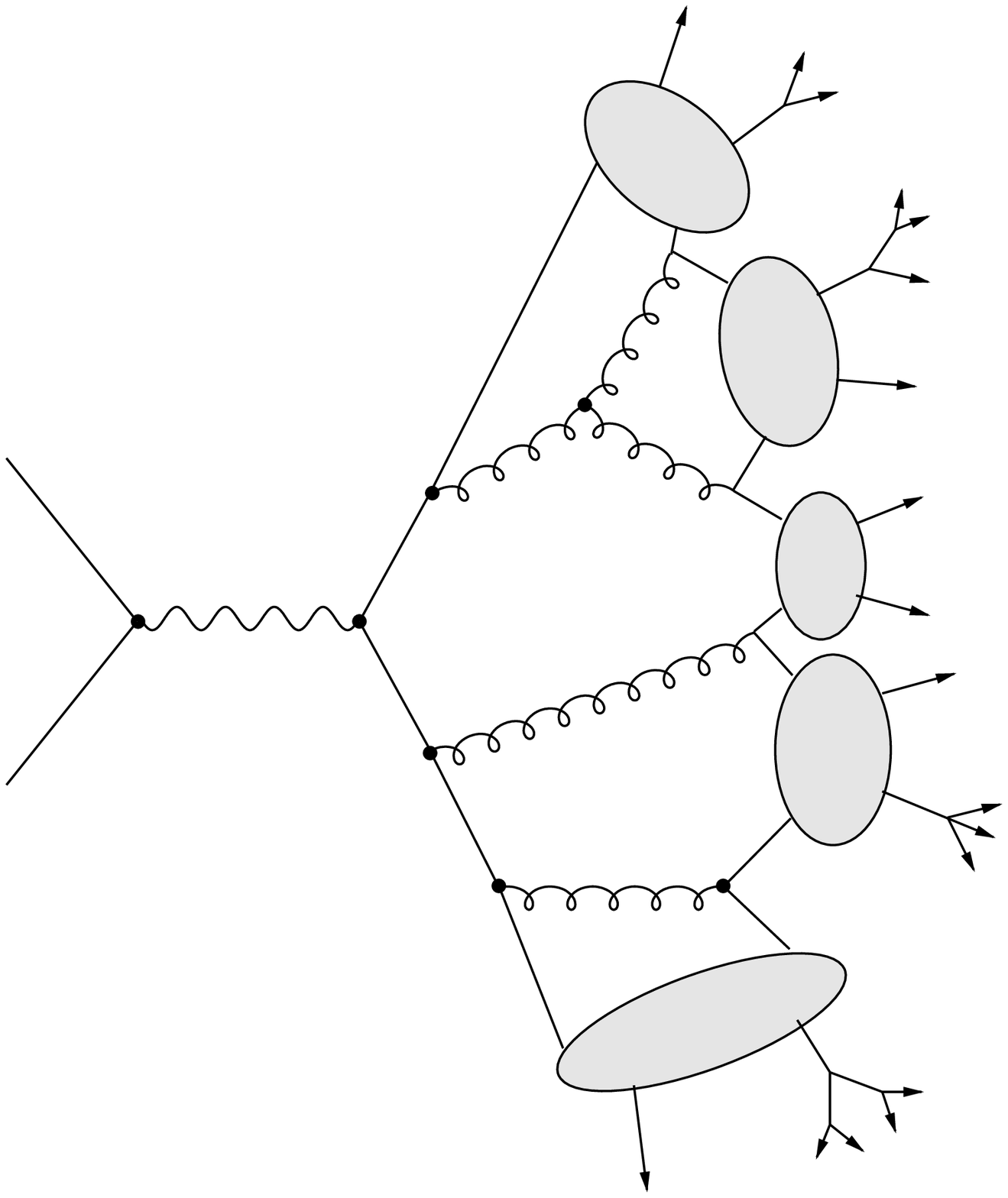,width=5cm}\end{minipage}
\begin{minipage}{50mm}
\epsfig{file=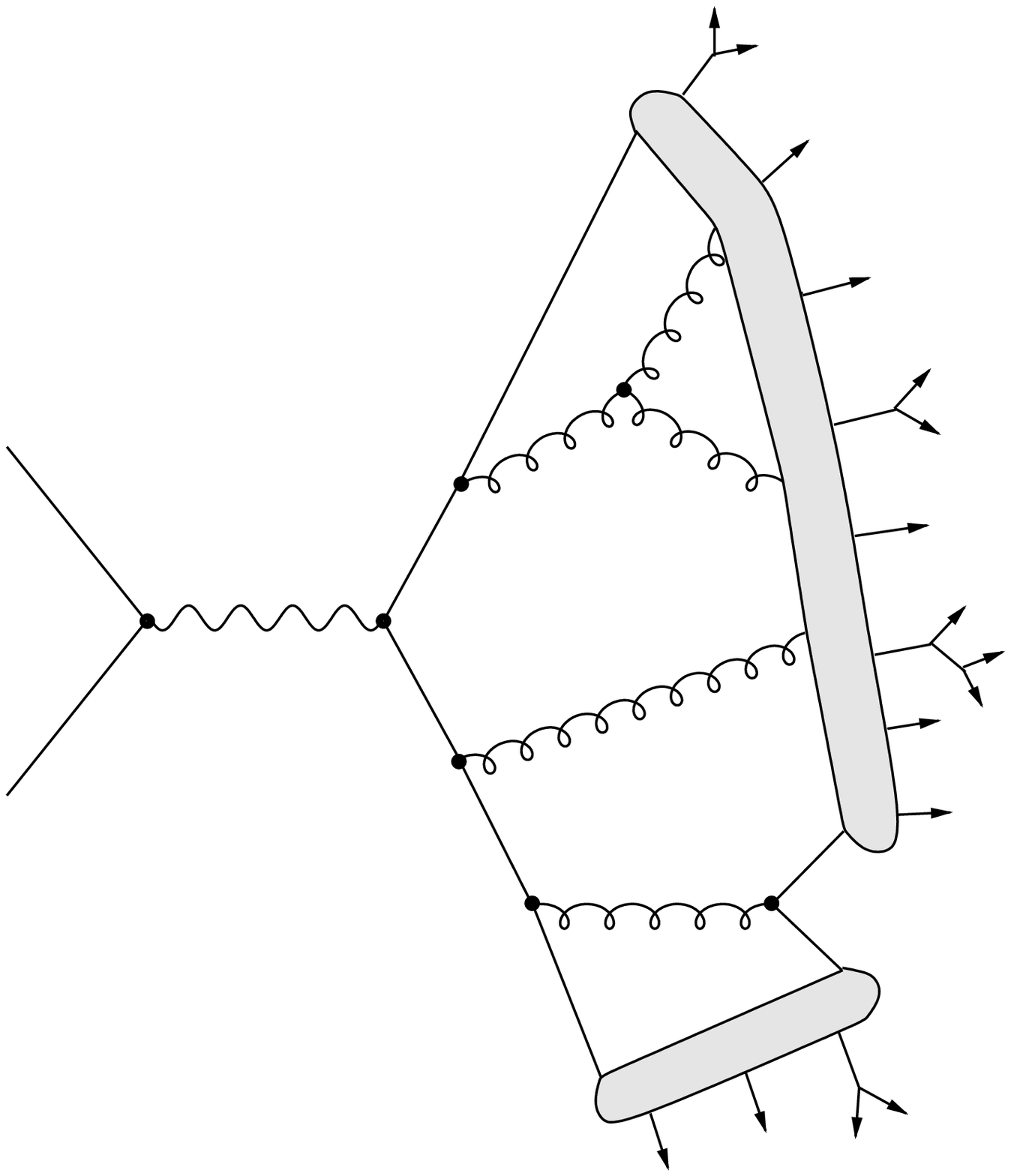,width=5cm}\end{minipage}
\caption{Cluster and string hadronization models.}
\label{fig:clus_string}\end{center}\end{figure}
\begin{itemize}
\item {\em Cluster model}
\cite{Marchesini:1984bm}-\cite{Marchesini:1996vc}.
  The model starts by splitting
  gluons non-perturbatively, $g\to q\bar q$, after the parton shower.
  Colour-singlet $q\bar q$ combinations have lower masses and a universal
  spectrum due to the {\em preconfinement}
  \cite{Amati:1979fg,Marchesini:1981cr} property of the shower
  (fig.~\ref{fig:kl_fig9} \cite{Knowles:1997dk}).
  These colour-singlet combinations are assumed to form clusters, which
  mostly undergo simple isotropic decay into pairs of hadrons, chosen
  according to the density of states with appropriate quantum numbers
  \cite{Webber:1984if}.
  This model has few parameters and a natural mechanism for generating
  transverse momenta and suppressing heavy particle production in
  hadronization. However, it has problems in dealing with the decay of
  very massive clusters, and in adequately suppressing baryon and heavy quark
  production.

\begin{figure}\begin{center}
\epsfig{file=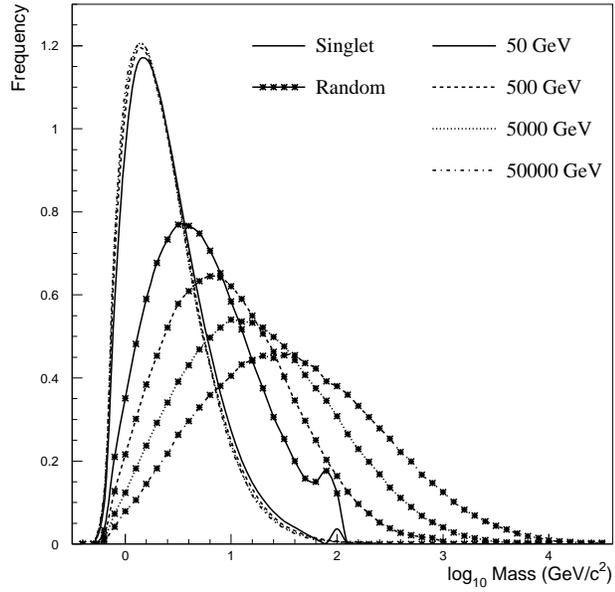,width=8cm}
\caption{Cluster model: mass distribution of $q\bar q$ pairs.}
\label{fig:kl_fig9}\end{center}\end{figure}
\begin{figure}\begin{center}
\epsfig{file=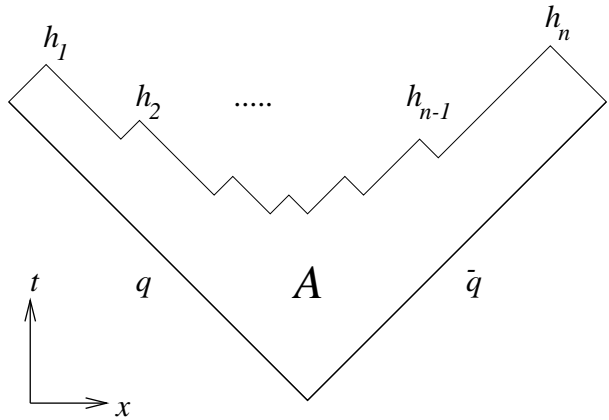,width=8cm}
\caption{String model: space-time picture.}
\label{fig:string_area}\end{center}\end{figure}

\item {\em String model}
\cite{Andersson:1983ia}-\cite{Sjostrand:1994yw}.
  This model is based on the dynamics of a
  relativistic string, representing the colour flux stretched between
  the initial $q\bar q$.  The string produces a linear confinement potential
  and an area law for matrix elements:
  $$ |M(q\bar q\to h_1\cdots h_n)|^2 \propto e^{-bA}$$
  where $A$ is the space-time area swept out
  (fig.~\ref{fig:string_area}).  The string breaks up into hadrons via
  $q\bar q$ pair production in its intense colour field.
  Gluons produced in the parton shower give rise to `kinks' on the string.
  The model has extra parameters for the transverse momentum distribution
  and heavy particle suppression. It has some problems describing baryon
  production, but less than the cluster model.

\item {\em The UCLA model} \cite{Buchanan:1987ua}-\cite{Chun:1997bh}
   is a variant of the JETSET string model which takes the above area law
   for matrix elements more seriously, using
   it to determine the relative rates of production of different hadron
   species. This results in heavy particle suppression without extra
   parameters, the mass-squared
   of a hadron being proportional to its space-time area. At present the
   model still uses extra parameters for $p_T$ spectra, and again has some
   problems describing baryon production.
\end{itemize}

\section{Single-particle yields and spectra}\label{sec:yields}
Tables \ref{tab:proc_mes} and
\ref{tab:proc_bar}\footnote{Updated from ref.~\cite{Knowles:1997dk}.}
compare predictions of the above models\footnote{Recent ALEPH HERWIG
tuning with strangeness suppression 0.8 \cite{Rudolph}.}
with data on Z$^0$ decay from LEP and SLC. Of course, the models have
tunable parameters, but the overall agreement is encouraging. As stated
earlier, the main problems are in the baryon sector, especially for
HERWIG.
\begin{table}\begin{center}{\small
\begin{tabular}{|c|c|c|c|c|c|} \hline 
Particle& Multiplicity& HERWIG& JETSET& UCLA& Expts \\
 & & 5.9 & 7.4 & 7.4 & \\
\hline
Charged       & 20.96(18) & 20.95 & 20.95 & 20.88 &  ADLMO \\
$\pi^\pm$     & 17.06(24) & 17.41 & 16.95 & 17.04 &  ADO \\
$\pi^0$       & 9.43(38)  & 9.97  & 9.59  & 9.61  &  ADLO \\
{\boldmath $\eta$}  & 0.99(4)   & 1.02  & 1.00  & \underline{0.78} & ALO \\
$\rho(770)^0$ & 1.24(10)  & 1.18  & 1.50  & 1.17 & AD \\
$\omega(782)$ & 1.09(9)  & 1.17  & 1.35  & 1.01 & ALO \\
{\boldmath $\eta'(958)$}&0.159(26)& 0.097 & 0.155 & 0.121 & ALO \\
{\boldmath f$_0(980)$}  &0.155(8) & \underline{0.111} &
$\sim$\underline{0.1} & --- & ADO \\
a$_0(980)^\pm$  & 0.14(6) & 0.240 & ---   & --- & O \\
$\phi(1020)$  & 0.097(7)  & 0.104 & \underline{0.194} &\underline{0.132}
& ADO \\
{\boldmath f$_2(1270)$} &0.188(14)& 0.186 &  $\sim 0.2$ & --- & ADO \\
f$_2'(1525)$  & 0.012(6)  & 0.021 & --- & --- & D \\
\hline
K$^\pm$       & 2.26(6)  & 2.16 & 2.30 & 2.24 & ADO \\
{\boldmath K$^0$}   & 2.074(14) & 2.05 & 2.07 & 2.06 & ADLO \\
K$^*(892)^\pm$& 0.718(44) & 0.670 & \underline{1.10} & 0.779 & ADO \\
K$^*(892)^0$  & 0.759(32) & 0.676 & \underline{1.10} & 0.760 & ADO \\
K$_2^*(1430)^0$ & 0.084(40) & 0.111 & --- & --- & DO \\
\hline
D$^\pm$         & 0.187(14) &  \underline{0.276} & 0.174 & 0.196 & ADO \\
D$^0$           & 0.462(26) & 0.506 & 0.490 & 0.497 & ADO \\
D$^*(2010)^\pm$ & 0.181(10) & 0.161 & \underline{0.242} & \underline{0.227}
& ADO \\
D$^\pm_{\rm s}$ & 0.131(20) & 0.115 & 0.129 & 0.130 & O \\
\hline
B$^*$           & 0.28(3)   & 0.201 & 0.260 & 0.254 & D \\
B$^{**}_{\rm u,d}$ & 0.118(24) & \underline{0.013} & ---   & --- & D \\
\hline
J/$\psi$           & 0.0054(4) & \underline{0.0018} & 0.0050 & 0.0050 & ADLO \\
$\psi(3685)$       & 0.0023(5) & 0.0009 & 0.0019 & 0.0019 & DO \\
$\chi_{{\mathrm c}1}$ & 0.0086(27) & \underline{0.0001} & --- & --- & DL \\
\hline
\end{tabular}}
\caption{Meson yields in Z$^0$ decay. Experiments: A=Aleph, D=Delphi,
L=L3, M=Mark II, O=Opal. Bold: new data this year. Underlined: disagreement
with data by more than 3$\sigma$.}
\label{tab:proc_mes}
\end{center}\end{table}
\begin{table}\begin{center}{\small
\begin{tabular}{|c|c|c|c|c|c|} \hline 
Particle& Multiplicity& HERWIG& JETSET& UCLA& Expts \\ 
 & & 5.9 & 7.4 & 7.4 &  \\
\hline
p     & 1.04(4) & \underline{0.863} & \underline{1.19} & 1.09 &  ADO \\
\hline
$\Delta^{++}$ & 0.079(15) & \underline{0.156} & \underline{0.189}
&\underline{0.139} & D \\
              & 0.22(6)   & 0.156 & 0.189 & 0.139 & O \\
\hline
{\boldmath $\Lambda$} & 0.399(8)& 0.387 & 0.385 & 0.382 & ADLO \\
{\boldmath$\Lambda(1520)$}  & 0.0229(25) & --- & --- & --- & DO \\
\hline
$\Sigma^\pm$  & 0.174(16) & 0.154 & 0.140 & 0.118 & DO \\
$\Sigma^0$    & 0.074(9) & 0.068  & 0.073 & 0.074 & ADO \\
$\Sigma^{\star\pm}$ & 0.0474(44) & \underline{0.111} & \underline{0.074}
& \underline{0.074} &
 ADO \\
\hline
$\Xi^-$       & 0.0265(9) & \underline{0.0493} & 0.0271 & \underline{0.0220}
& ADO \\
$\Xi(1530)^0$ & 0.0058(10) & \underline{0.0205} & 0.0053 & 0.0081 & ADO \\
\hline
$\Omega^-$    & 0.0012(2) & \underline{0.0056} & 0.00072 & 0.0011 & ADO \\
\hline
$\Lambda_{\rm c}^+$ & 0.078(17) & \underline{0.0123}  & 0.059 &
\underline{0.026} & O \\
\hline
\end{tabular}}
\caption{Baryon yields in Z$^0$ decay. Legend as in table 1.}
\label{tab:proc_bar}
\end{center}\end{table}

\begin{figure}\begin{center}
\epsfig{file=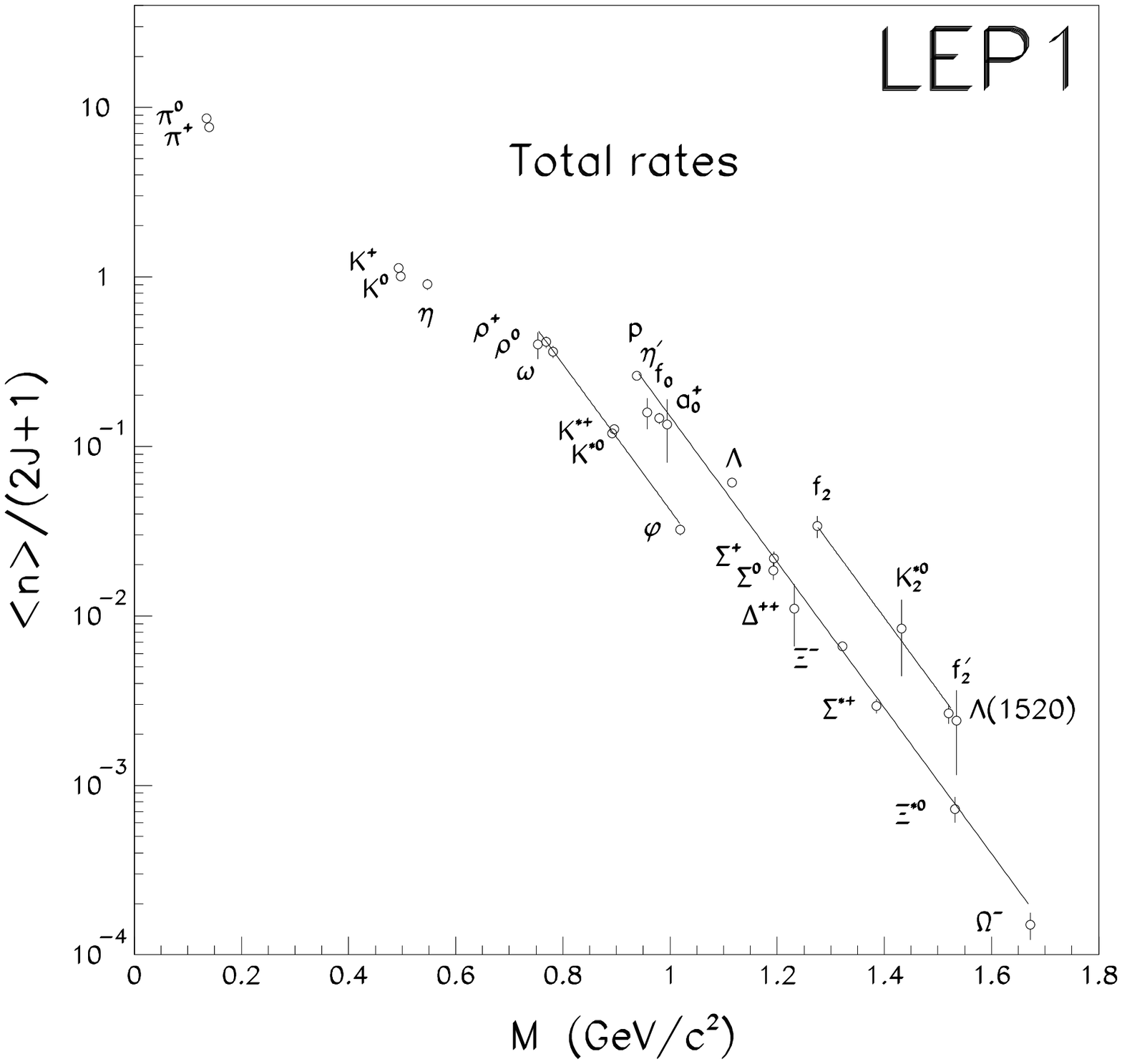,width=8cm}
\caption{Particle yields in Z$^0$ decay.}
\label{fig:chliap}\end{center}\end{figure}

It is remarkable that most measured yields (except for
the $0^-$ mesons, which have special status as Goldstone
bosons) lie on the family of curves
$$\VEV{n} = a (2J+1) e^{-M/T}$$
where $M$ is the mass and $T\simeq 100$ MeV
(fig.~\ref{fig:chliap} \cite{Chliapnikov:1999qi}).
This suggests that mass, rather than quantum numbers, is
the primary factor in determining production rates. Note that,
surprisingly, the orbitally-excited $J=\thrhf$ baryon
$\Lambda(1520)$ (not yet included in models) is produced
almost as much as the unexcited $J=\thrhf$ baryon $\Sigma(1385)$
\cite{Alexander:1996qj,3_147}.

\begin{figure}\begin{center}
\epsfig{file=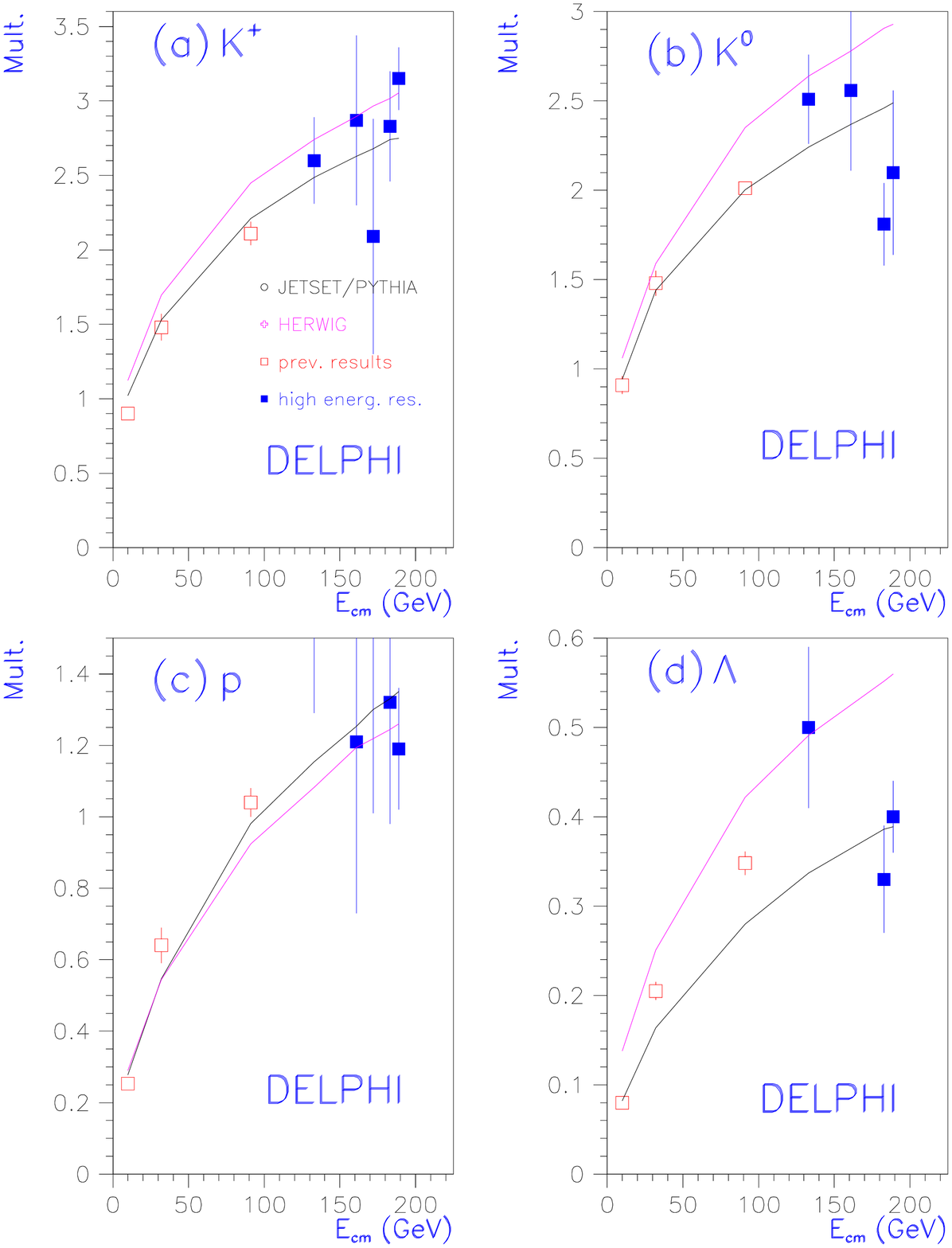,width=8cm}
\caption{Particle yields in $\ee$ annihilation.}
\label{fig:1_229_16}\end{center}\end{figure}

\begin{figure}\begin{center}
\begin{minipage}{60mm}
\epsfig{file=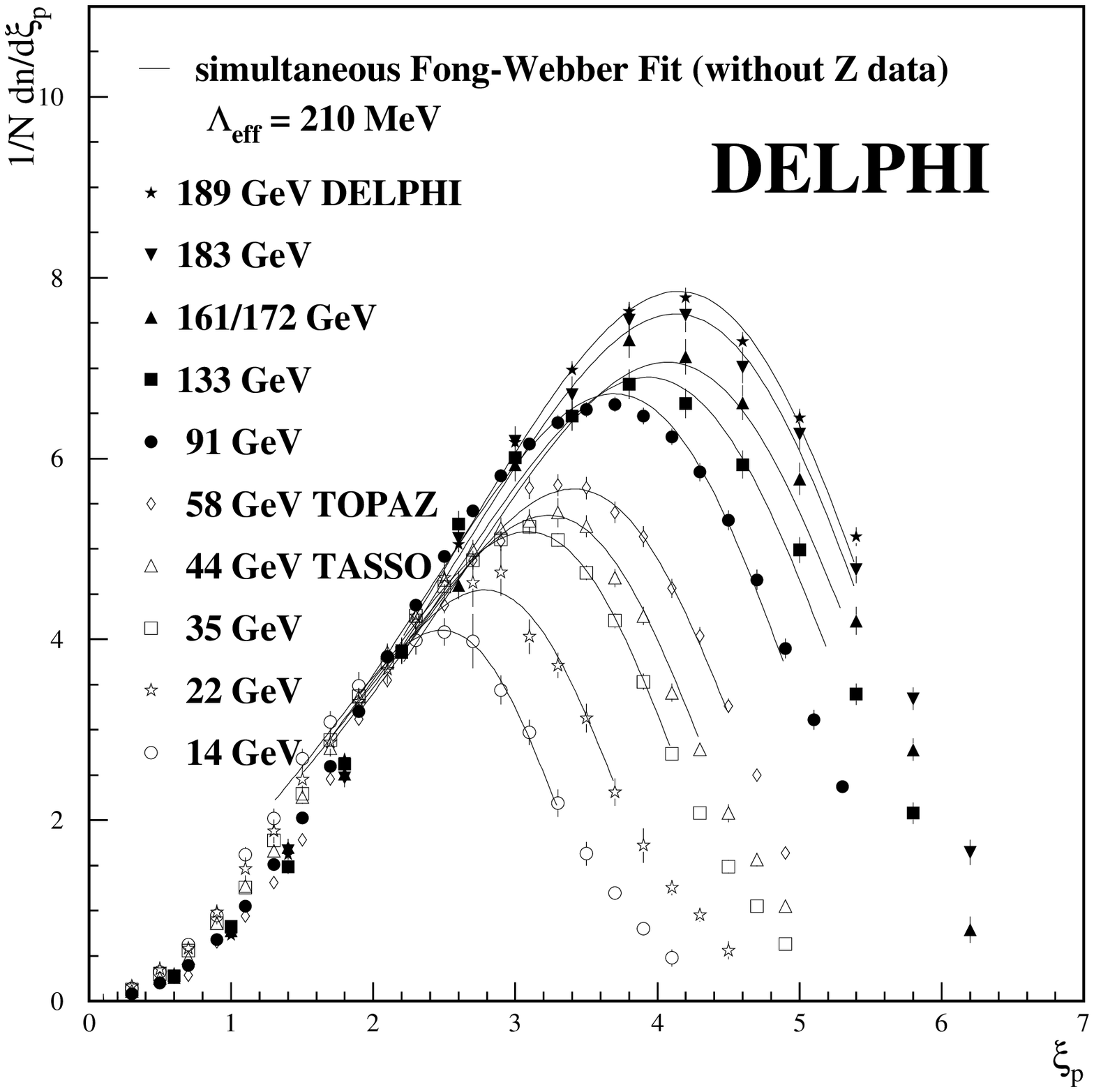,width=60mm}
\end{minipage}
\begin{minipage}{60mm}
\epsfig{file=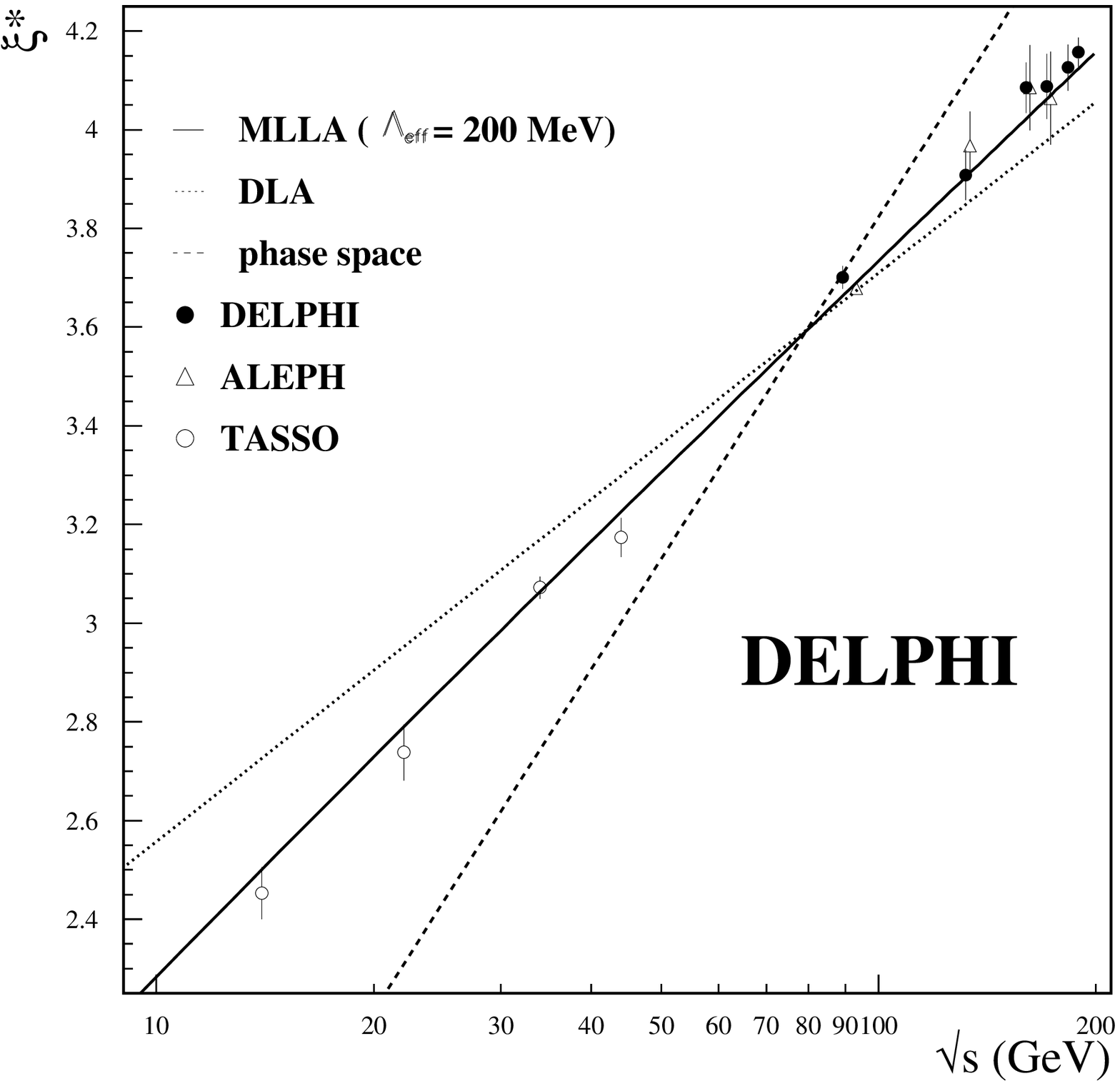,width=60mm}
\end{minipage}
\caption{Low-$x$ fragmentation in $\ee$ annihilation.}
\label{fig:1_225_2}\end{center}\end{figure}

At other energies, model predictions for identified particle yields
are in broad agreement with $\ee$ data (fig.~\ref{fig:1_229_16} \cite{1_229}),
but statistics are of course poorer. Charged particle spectra at low $x$
agree well with the resummed (MLLA) predictions 
\cite{Azimov:1986by,Fong:1989qy,Dokshitzer:1991wu}
over a wide  energy range, as illustrated
in fig.~\ref{fig:1_225_2} \cite{1_225}.

\begin{figure}\begin{center}
\epsfig{file=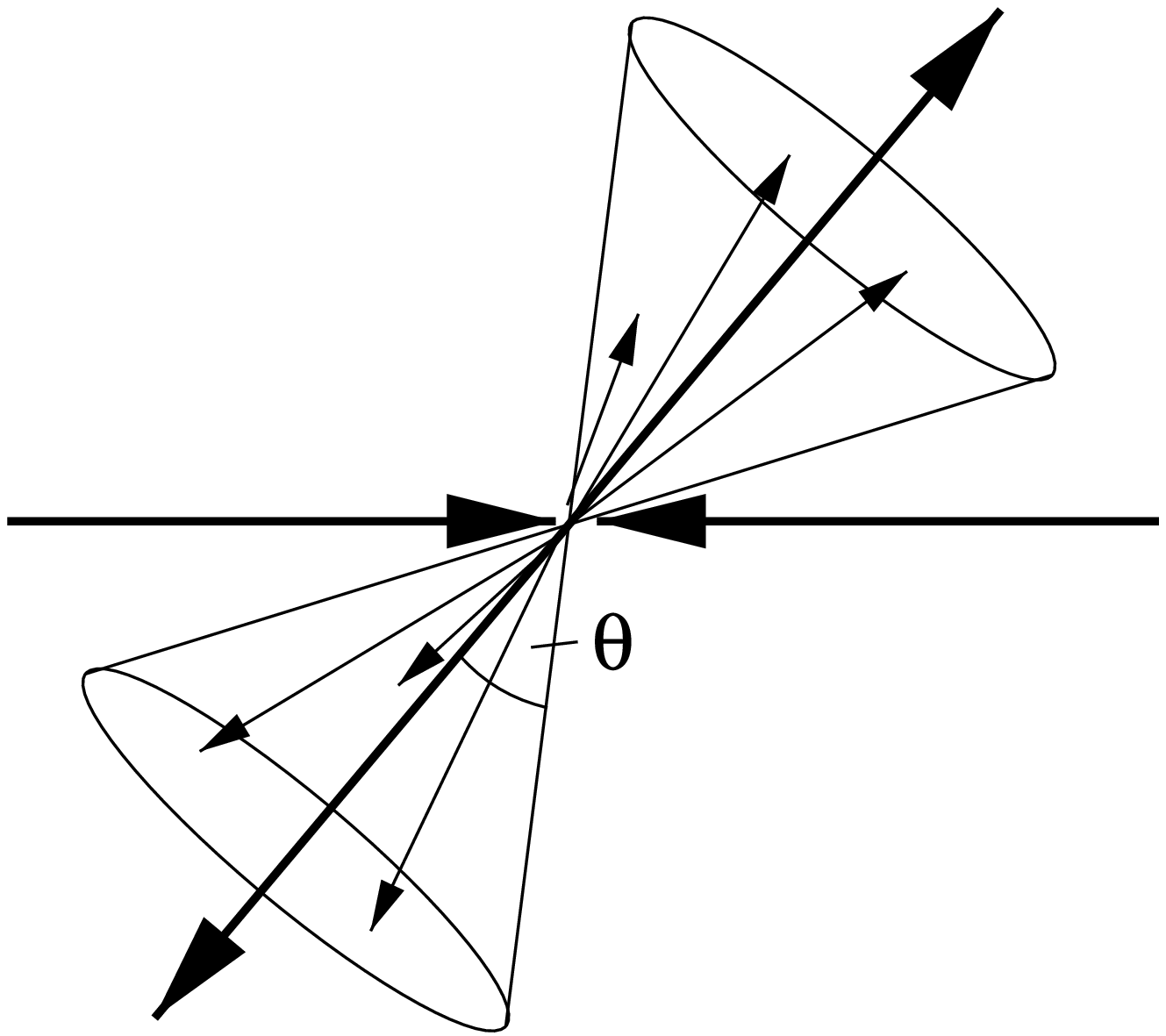,width=60mm}
\caption{Cone angle in  $p\bar p \to$ dijets}
\label{fig:dijet}\end{center}\end{figure}

\begin{figure}\begin{center}
\begin{minipage}{70mm}
\epsfig{file=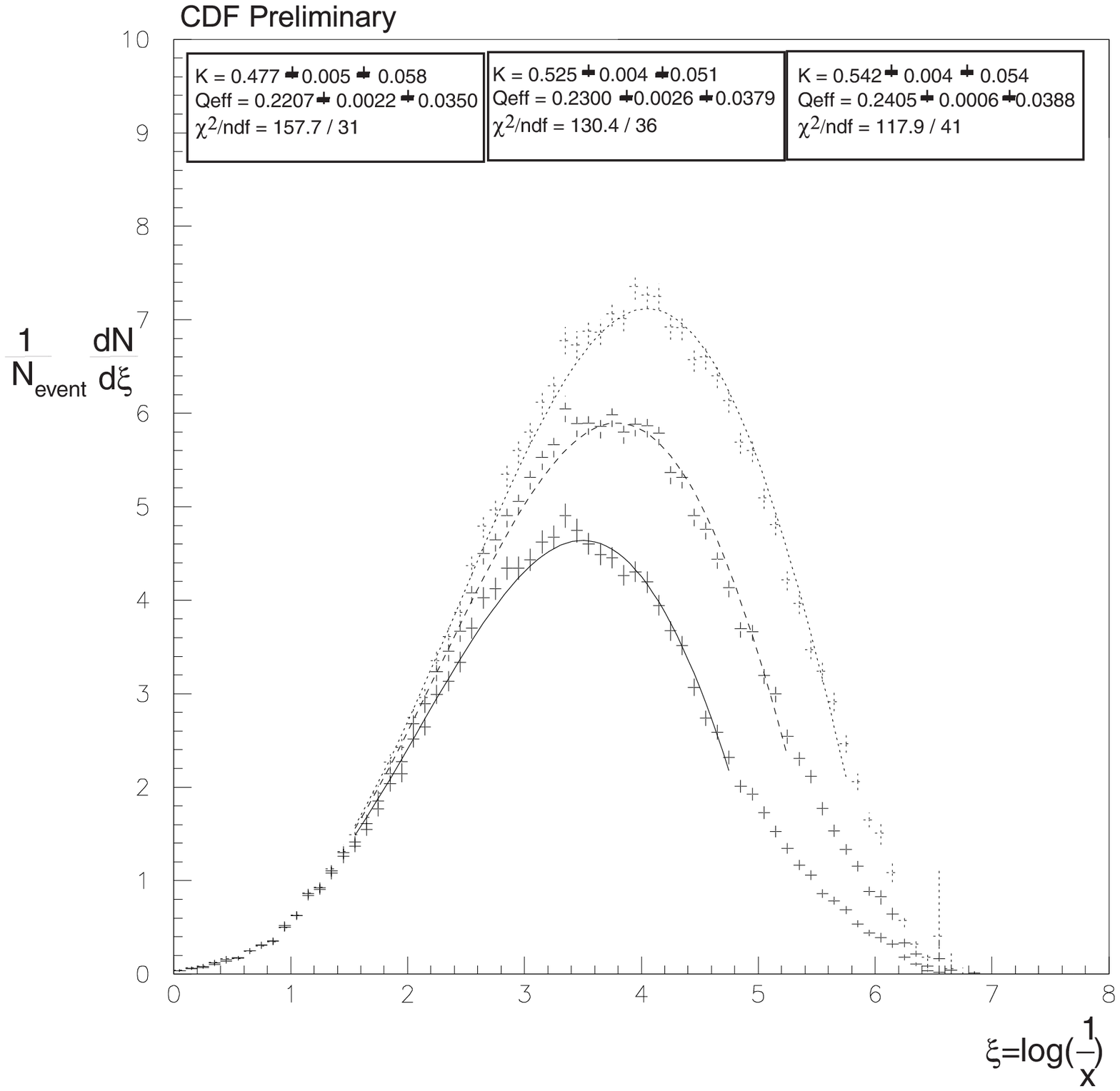,width=65mm}
\end{minipage}
\begin{minipage}{70mm}
\epsfig{file=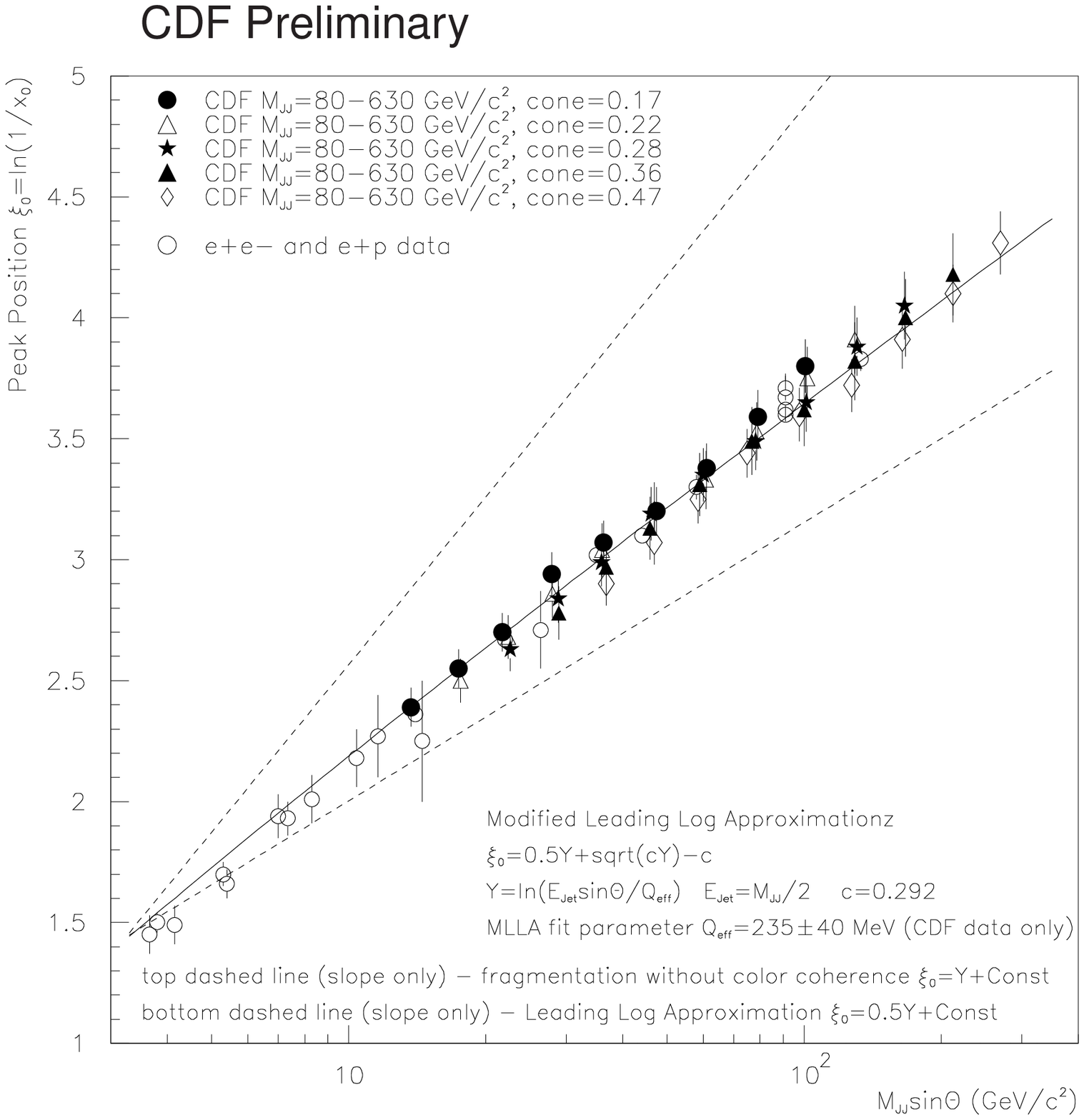,width=65mm}
\end{minipage}
\caption{Low-$x$ fragmentation in $p\bar p \to$ dijets.}
\label{fig:cdf4996_peak_vs_m_new}\end{center}\end{figure}

In $p\bar p \to$ dijets \cite{cdf4996} the relevant scale is taken to be
{$Q=M_{JJ}\sin\theta$} where $M_{JJ}$ is the dijet mass and
$\theta$ is the jet cone angle (fig.~\ref{fig:dijet}). Results
are then in striking agreement with theory and with
data from $\ee$ annihilation at $Q=\sqrt s$
(fig.~\ref{fig:cdf4996_peak_vs_m_new}).

\begin{figure}\begin{center}
\epsfig{file=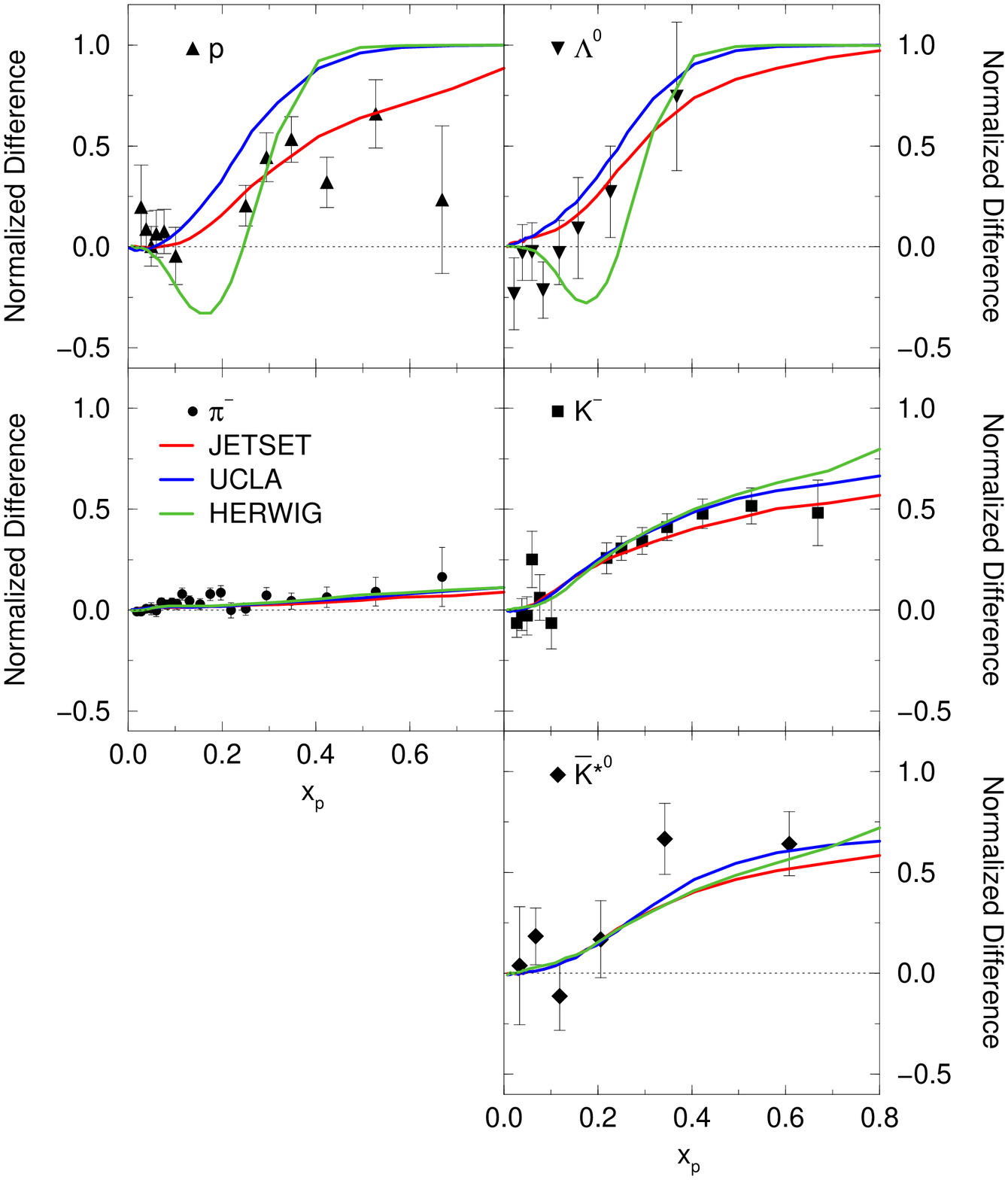,width=70mm}
\caption{Normalized particle--antiparticle differences
in quark jet fragmentation.}
\label{fig:SLD_8159_8}\end{center}\end{figure}

New SLD data include hadron spectra in light quark (rather than antiquark)
fragmentation, selected by hemisphere using the SLC beam polarization
\cite{Abe:1999qh}. One sees strong particle/antiparticle differences
in the expected directions (fig.~\ref{fig:SLD_8159_8}),
bearing in mind the predominance of down-type quarks in Z$^0$ decay.
\begin{figure}\begin{center}
\epsfig{file=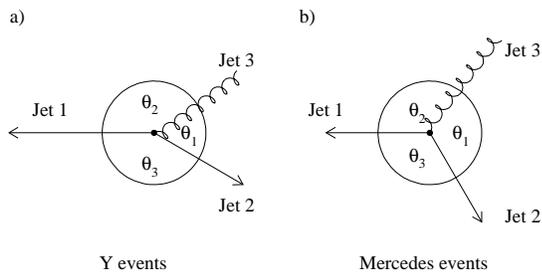,width=80mm}
\caption{Selection of gluon jets by DELPHI}
\label{fig:Delphi_3_146_1}\end{center}\end{figure}

\section{Quark and gluon jets}\label{sec:qg}
\begin{figure}\begin{center}\epsfig{file=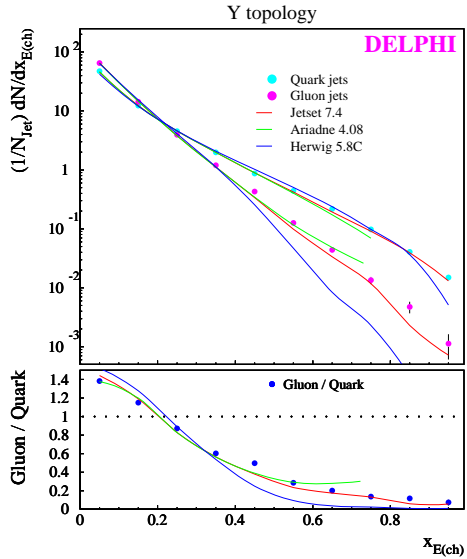,width=70mm}
\caption{Charged particle spectra in quark and gluon jets.}
\label{fig:Delphi_1_571_9}\end{center}\end{figure}

\begin{figure}\begin{center}\epsfig{file=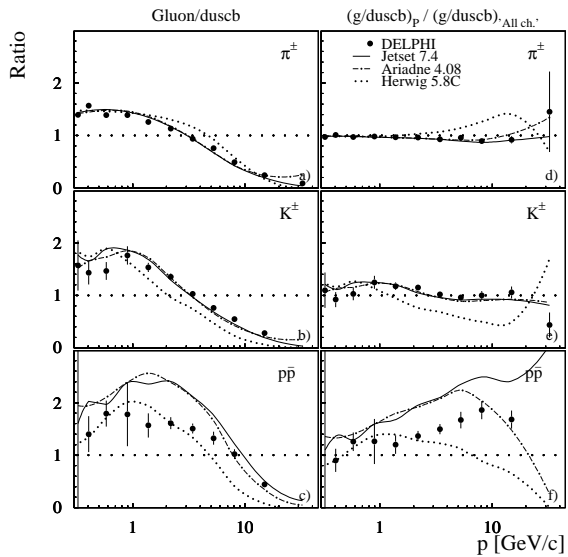,width=80mm}
\caption{Comparisons of particle spectra in quark and gluon jets.}
\label{fig:Delphi_3_146_53}\end{center}\end{figure}

DELPHI \cite{3_146} select gluon jets by anti-tagging
heavy quark jets in Y and Mercedes three-jet events
(fig.~\ref{fig:Delphi_3_146_1}).  As expected, the higher
colour charge of the gluon ($C_A=3$ vs.\ $C_F=4/3$) leads to
a softer spectrum and higher overall multiplicity
(fig.~\ref{fig:Delphi_1_571_9}).
In general the relative multiplicities of identified particles are
consistent with those of all charged, with no clear excess of
any species in gluon jets (fig.~\ref{fig:Delphi_3_146_53}).
In particular there is no enhanced $\phi(1020)$ or $\eta$ production:

\noindent
DELPHI \cite{3_146}: $N_g(\phi)/N_q(\phi) = 0.7\pm 0.3$\\
OPAL \cite{1_4}: $N_g(\eta)/N_q(\eta) = 1.29\pm 0.11$


\begin{figure}\begin{center}
\begin{minipage}{70mm}
\epsfig{file=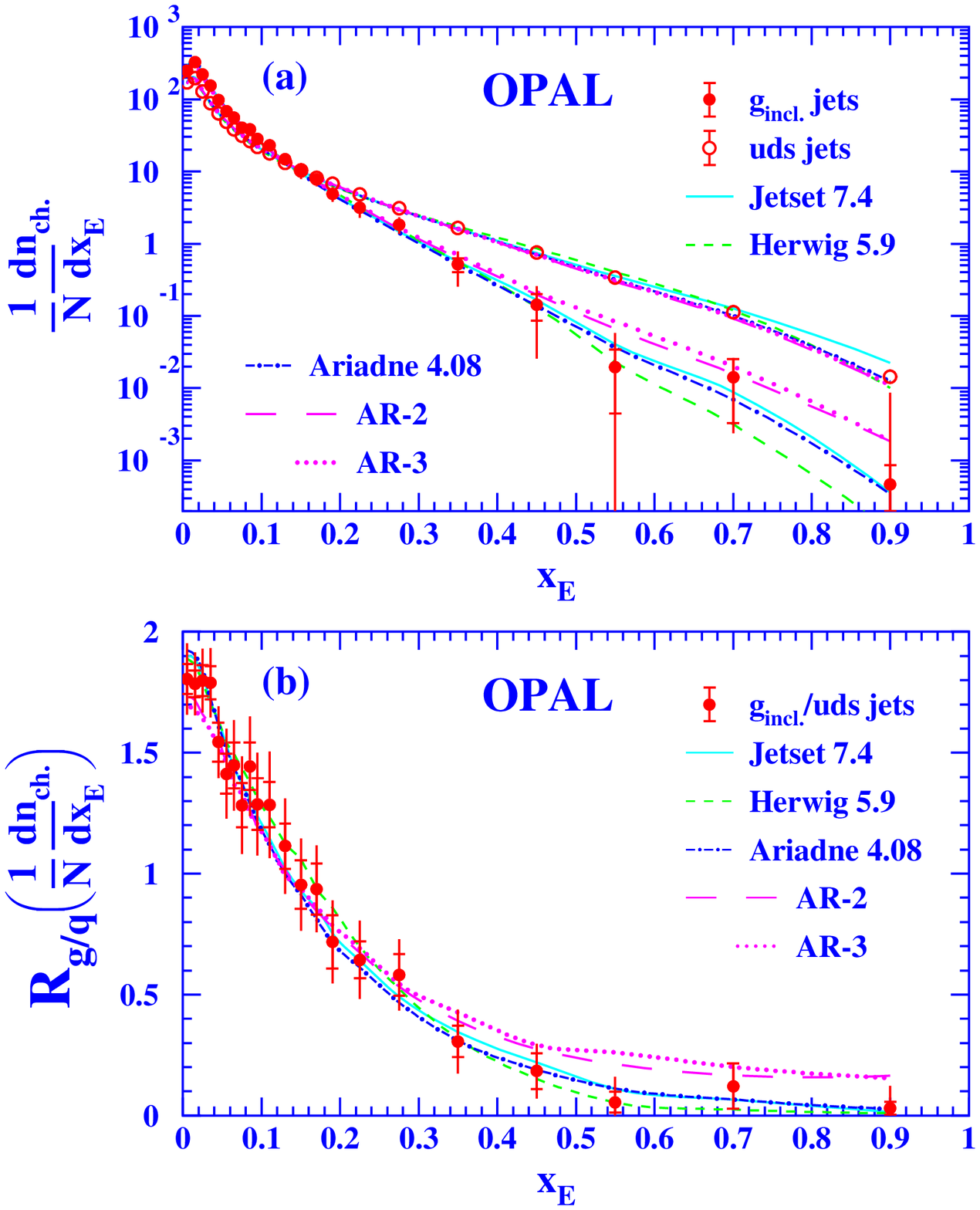,width=70mm}
\end{minipage}
\begin{minipage}{70mm}
\epsfig{file=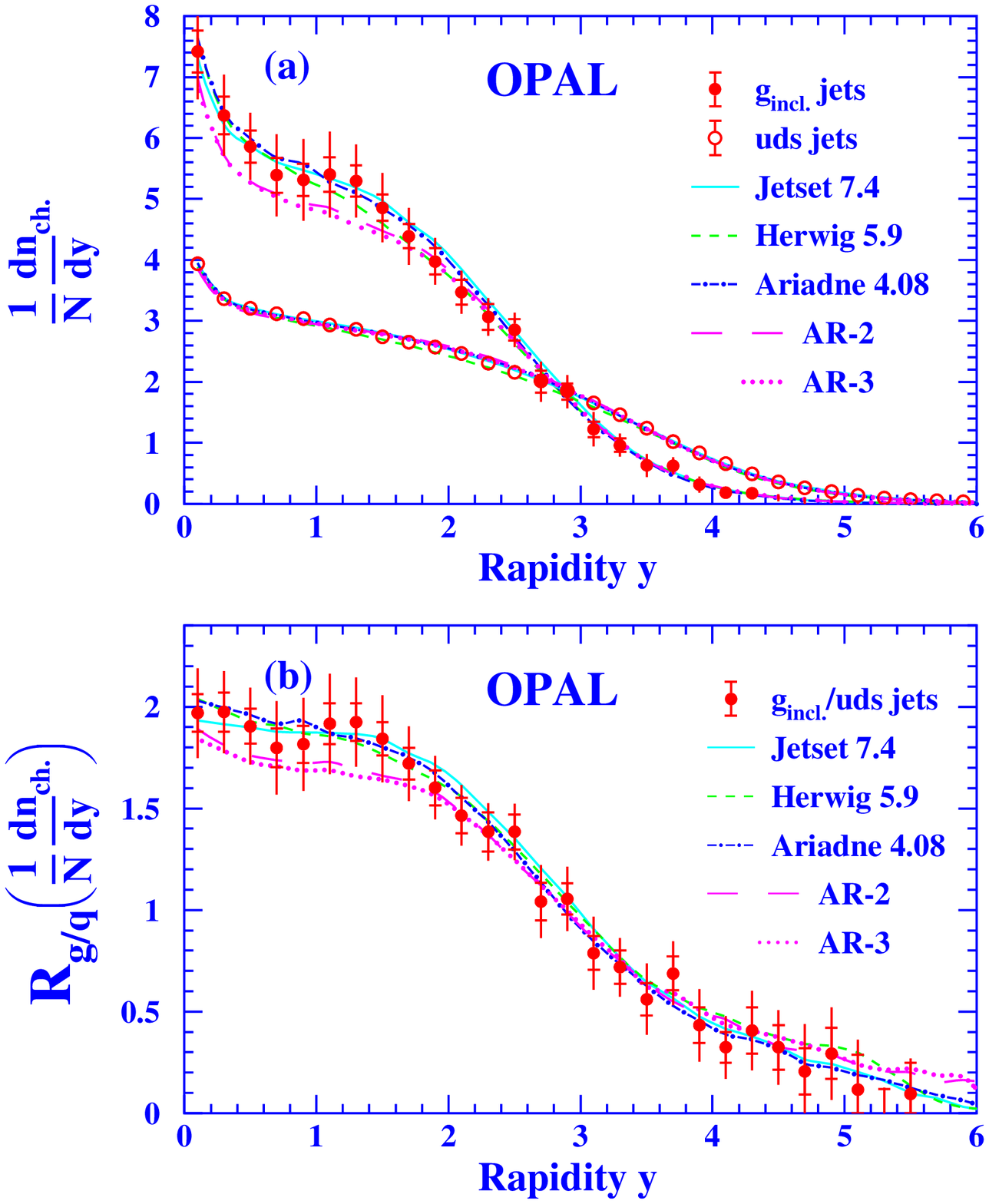,width=70mm}
\end{minipage}
\caption{Momentum fraction and rapidity distributions in quark
and gluon jets.}\label{fig:Opal_24_5}\end{center}\end{figure}

OPAL \cite{Abbiendi:1999pi} select gluon jets recoiling against two
tagged $b$-jets in the same hemisphere. Monte Carlo studies indicate
that such jets should be similar to those emitted by a point source of
gluon pairs. The qualitative message from the data is again clear
(fig.~\ref{fig:Opal_24_5}): Gluon jets have softer fragmentation
than light quark jets, and higher multiplicity.
The precision of the data is now such that
next-to-leading order calculations of the relevant
coefficient functions, taking into account the experimental
selection procedures, are needed to check universality
of the extracted gluon fragmentation function.

The ratio of gluon/quark multiplicities at {low rapidity} (large
angle) is close to the ratio of colour charges
{$r\equiv C_A/C_F=2.25$},
in agreement with local parton-hadron duality:
$$\mbox{OPAL: }
{r_{ch}(|y|<1) = 1.919\pm 0.047\pm 0.095}$$

Monte Carlo studies \cite{Abbiendi:1999pi}
suggest that a better measure of $C_A/C_F$ 
is obtained by selecting low-momentum hadrons with relatively
large $p_T$ (i.e.\ low rapidity).  This gives
$$\mbox{OPAL: }
{r_{ch}(p<4,\,0.8<p_T<3\;\mbox{GeV}) = 2.29\pm 0.09\pm 0.015}$$

\begin{figure}\begin{center}
\begin{minipage}{70mm}
\epsfig{file=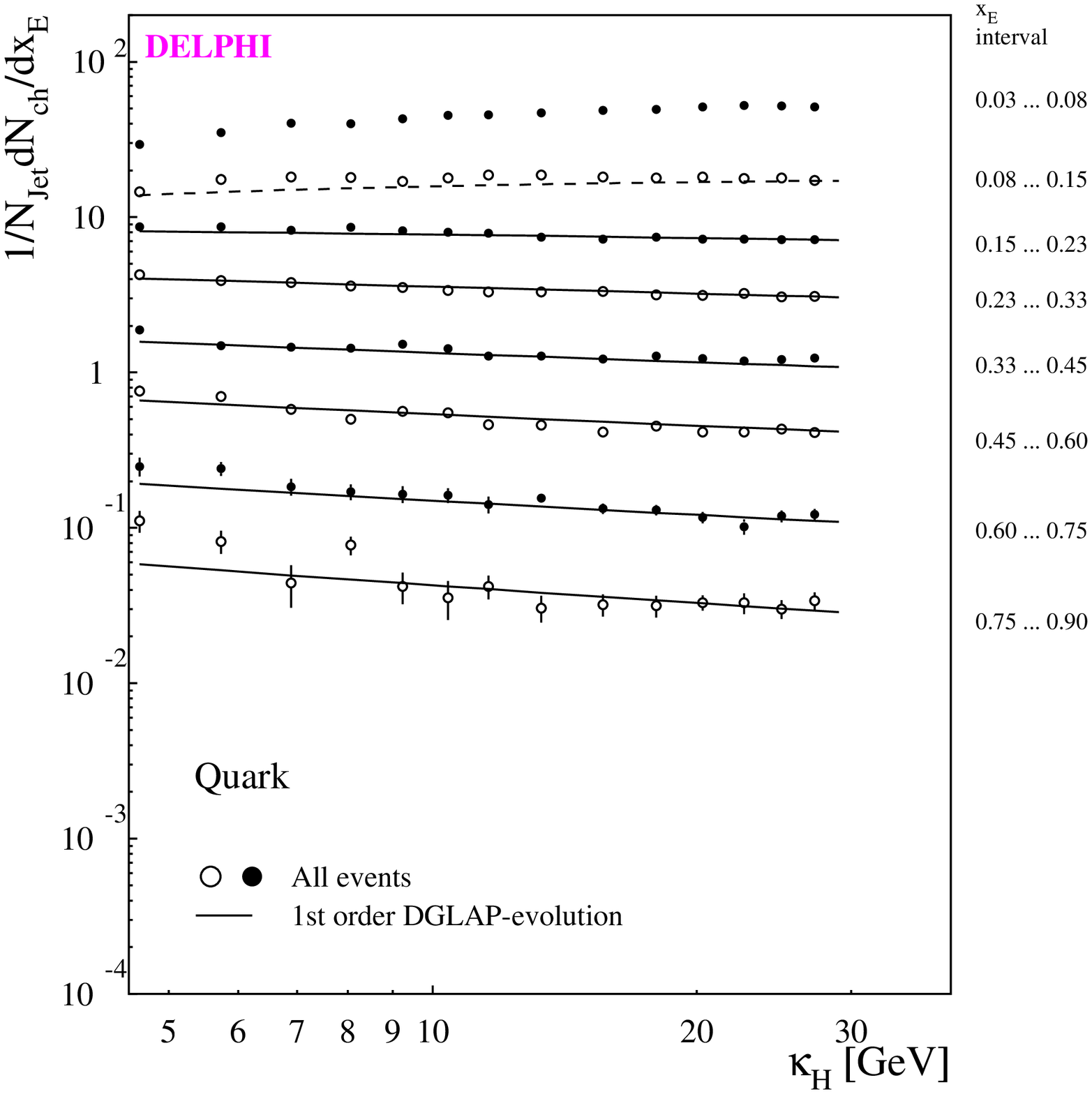,width=65mm}\end{minipage}
\begin{minipage}{70mm}
\epsfig{file=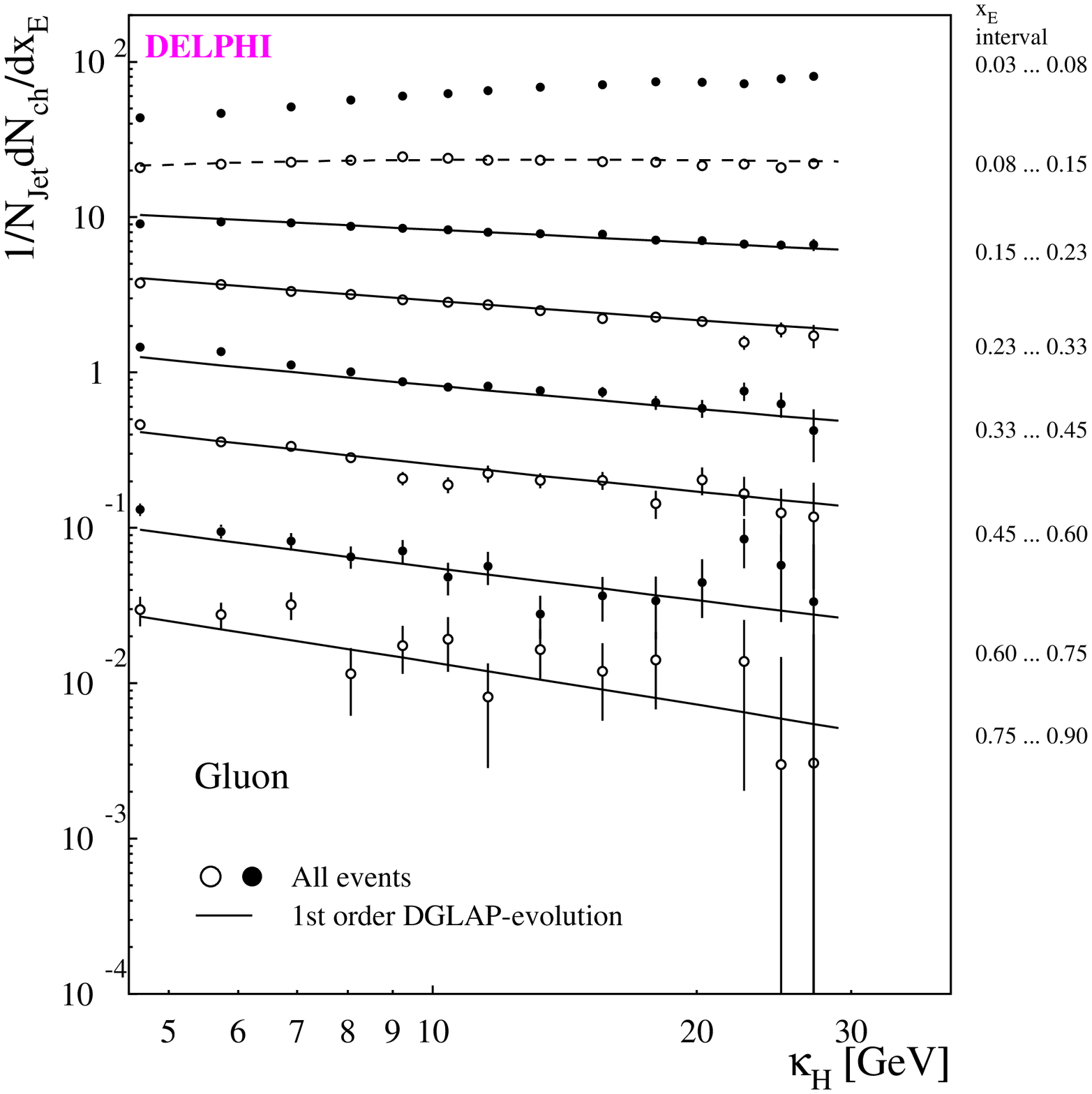,width=65mm}\end{minipage}
\caption{Scale dependence of quark and gluon fragmentation.}
\label{fig:Delphi_1_571_16}\end{center}\end{figure}

\begin{figure}\begin{center}\epsfig{file=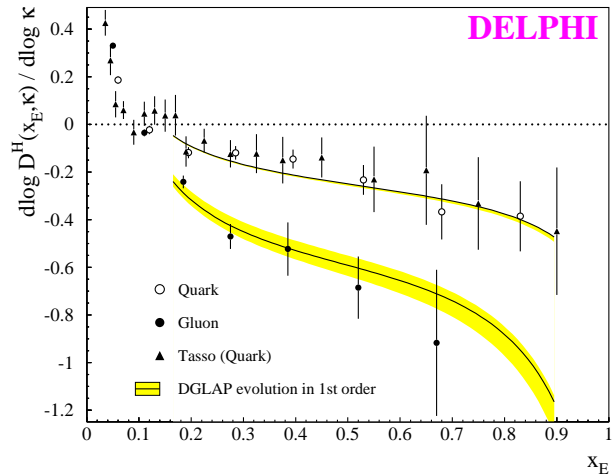,width=85mm}
\caption{Logarithmic gradients of quark and gluon fragmentation.}
\label{fig:Delphi_1_571_17}\end{center}\end{figure}

DELPHI \cite{1_571} have observed scaling violation in quark and gluon jet
fragmentation separately (fig.~\ref{fig:Delphi_1_571_16})
by studying the dependence on the scale
$$ \kappa_H = E_{jet}\,\sin (\theta/2)\;\simeq\;\half\sqrt{sy_3}$$
where $\theta$ is the angle to the closest jet and
$y_3$ is the Durham jet resolution \cite{Catani:1991hj}
at which 3 jets are just resolved.  This is expected
to be the relevant scale when $y_3$ becomes small.
One sees clearly that there is more scaling violation in gluon jets
(fig.~\ref{fig:Delphi_1_571_17}). The ratio provides another
measure of $C_A/C_F$:
$$\mbox{DELPHI: }
{r_{\mbox{\scriptsize sc.viol.}} = 2.23\pm 0.09\pm 0.06}$$


A crucial point in the DELPHI analysis is that 3-jet events are not
selected using a fixed jet resolution {$\ycut$}, but rather
each event is clustered to precisely 3 jets.  This avoids `biasing'
the gluon jet sample by preventing further jet emission above $\ycut$.

\begin{figure}\begin{center}\epsfig{file=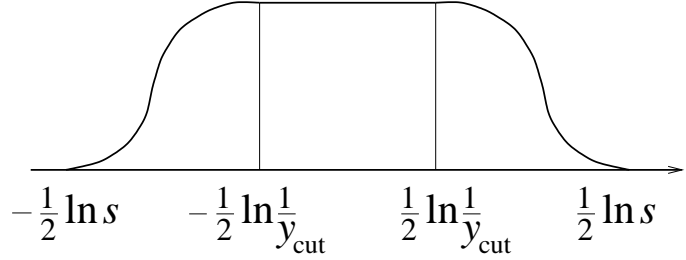,width=90mm}
\caption{Rapidity plateau in two-jet events.}
\label{fig:twojet}\end{center}\end{figure}

\begin{figure}\begin{center}\epsfig{file=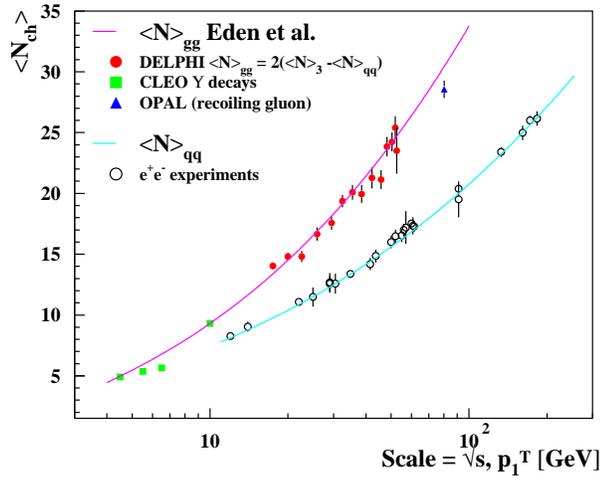,width=80mm}
\caption{Average $q\bar q$ and $gg$ multiplicities deduced from 2- and 3-jet
events.}\label{fig:Delphi_extra_ngg}\end{center}\end{figure}

The same point is well illustrated in analyses of average
multiplicities in 2- and 3-jet events
\cite{Catani:1992tm,Eden:1999vc,Abreu:1999rs}.
If $N_{q\bar q}(s)$ is the `unbiased' $q\bar q$ multiplicity,
then in events with precisely 2 jets at resolution $\ycut$ there
a rapidity plateau of length $\ln(1/\ycut)$ (see fig.~\ref{fig:twojet})
and the multiplicity is
$$ N_2(s,\ycut) \simeq N_{q\bar q}(s\ycut)
+\ln(1/\ycut)N'_{q\bar q}(s\ycut)$$
where $N'(s)\equiv sdN/ds$.  Clustering each event to 3 jets we get this
multiplicity with {$y_3$} in place of $\ycut$, plus an unbiased gluon jet:
{$$ N_3(s) \simeq N_2(s,y_3) +\half N_{gg}(sy_3)$$}
Thus one can extract the unbiased $gg$ multiplicity,
plotted in fig.~\ref{fig:Delphi_extra_ngg}
vs. $p_1^T\sim \sqrt{sy_3}$ \cite{Delphi_extra}.
The ratio of $gg/q\bar q$ slopes gives yet another measure of $C_A/C_F$
\cite{Abreu:1999rs}:
$$r_{\mbox{\scriptsize mult}} = 2.246\pm 0.062(\mbox{stat.})
\pm 0.080(\mbox{sys.})\pm 0.095(\mbox{theo.})$$


\section{Current and target fragmentation in DIS}\label{sec:DIS}
\begin{figure}\begin{center}\epsfig{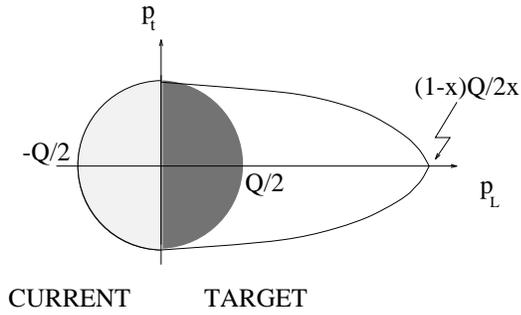}
\caption{Breit frame current and target regions in DIS.}
\label{fig:Zeus_537_1}\end{center}\end{figure}

H1 \cite{Adloff:1997fr} and ZEUS \cite{Breitweg:1999nt} have studied
the distributions of $x_p=2|\bom p|/Q$
in the current and target hemispheres in the Breit frame
(fig.~\ref{fig:Zeus_537_1}).

\begin{figure}\begin{center}\epsfig{file=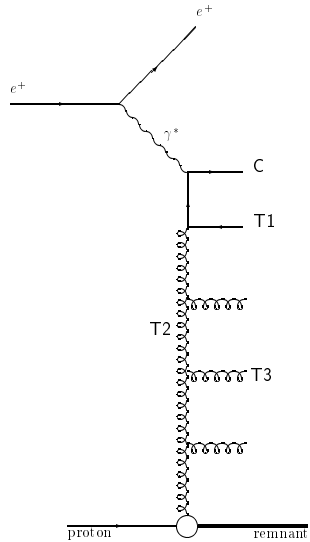,width=40mm}
\caption{Contributions to the final state in DIS.}
\label{fig:Zeus_537_2}\end{center}\end{figure}

In the current hemisphere one expects fragmentation of the current
jet (C in fig.~\ref{fig:Zeus_537_2}), similar to half an $\ee$ event.
In the target hemisphere, the contribution T1 is
similar to C, T2 gives extra particles with $x_p<1$, while T3 gives
$x_p\gtap 1$, generally outside detector acceptance.
\begin{itemize}
\item In the {\em current hemisphere} the charged multiplicity is indeed
similar to $\ee$ (fig.~\ref{fig:Zeus_537_10fit} \cite{Breitweg:1999nt}).
Differences at low $Q^2$ are consistent with the expected
boson-gluon fusion contribution.
The distribution of $\xi=\ln(1/x_p)$ is also
similar to $\ee$, i.e.\ close to Gaussian with little Bjorken $x$
dependence (fig.~\ref{fig:Zeus_537_12}).

\begin{figure}\begin{center}\epsfig{file=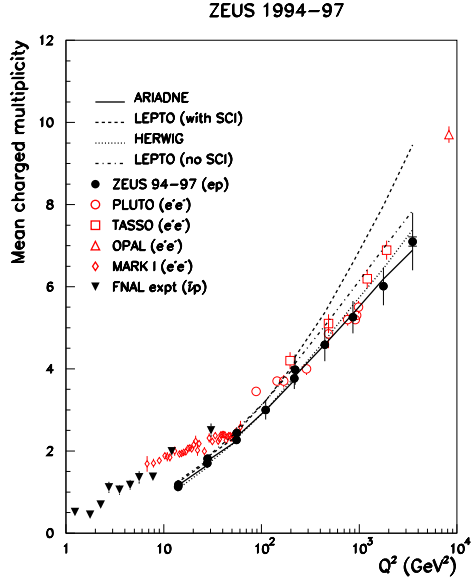,width=8cm}
\caption{Charged multiplicity in current hemisphere.}
\label{fig:Zeus_537_10fit}\end{center}\end{figure}
\begin{figure}\begin{center}\epsfig{file=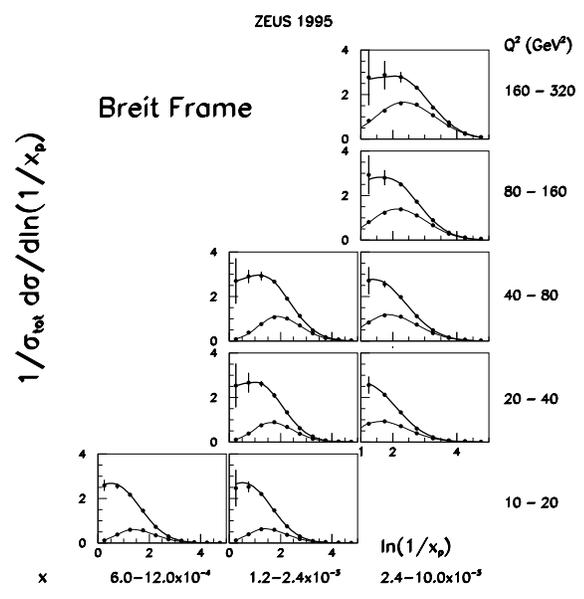,width=80mm}
\caption{Fragmentation in DIS. Upper data (heavy curve) target region,
lower data (light curve) current region.}\label{fig:Zeus_537_12}
\end{center}\end{figure}

At low $Q^2$ there is evidence of strong subleading corrections.
The distribution is skewed towards higher values of
$\xi$ (smaller $x_p$), contrary to MLLA predictions
(fig.~\ref{fig:Zeus_537_4sk}). The quantity plotted is
$$\mbox{Skewness} \equiv
\VEV{(\xi-\bar\xi)^3}/\VEV{(\xi-\bar\xi)^2}^{\thrhf}$$

\begin{figure}\begin{center}\epsfig{file=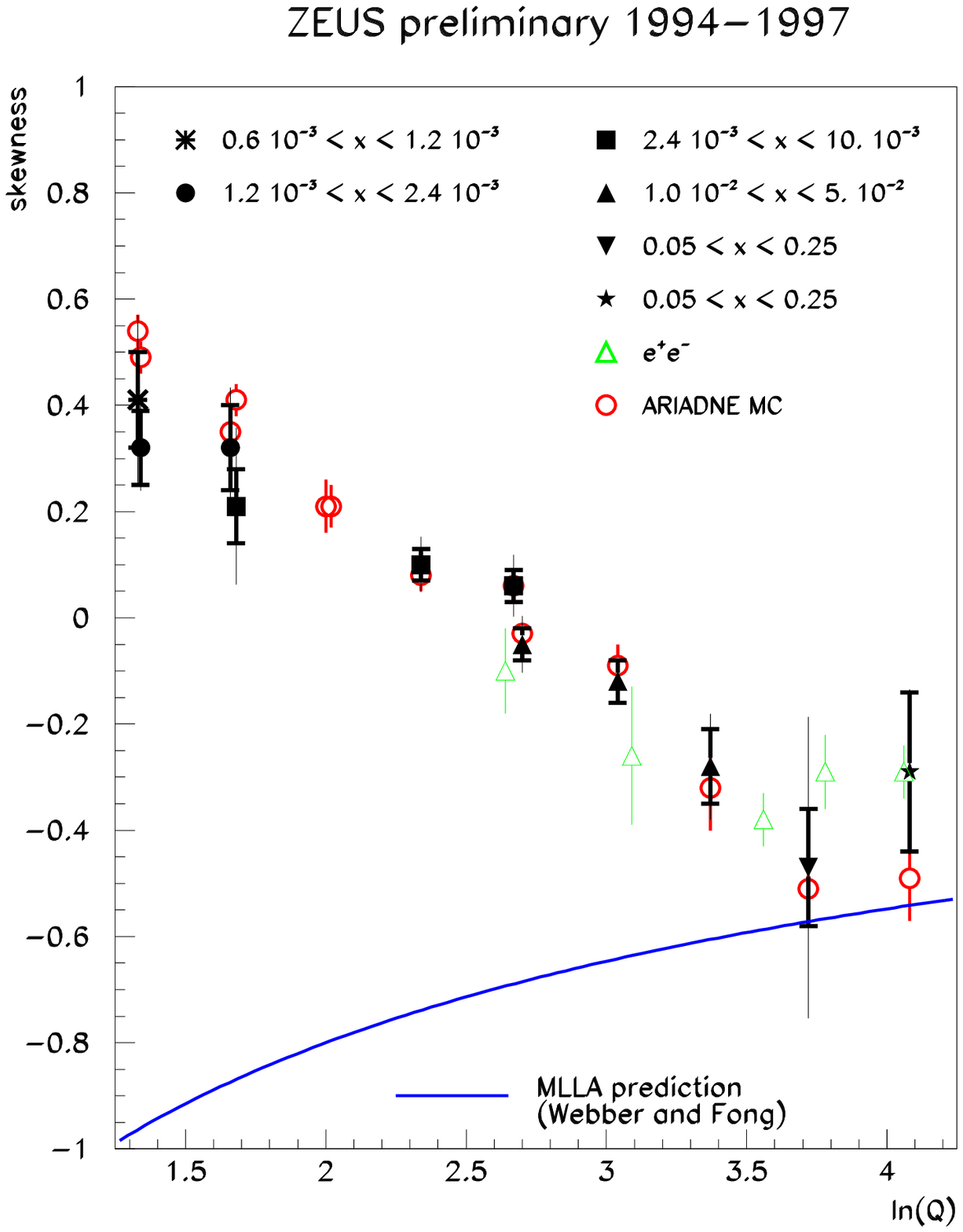,width=60mm}
\caption{Skewness in current fragmentation region.}
\label{fig:Zeus_537_4sk}\end{center}\end{figure}
\begin{figure}\begin{center}\epsfig{file=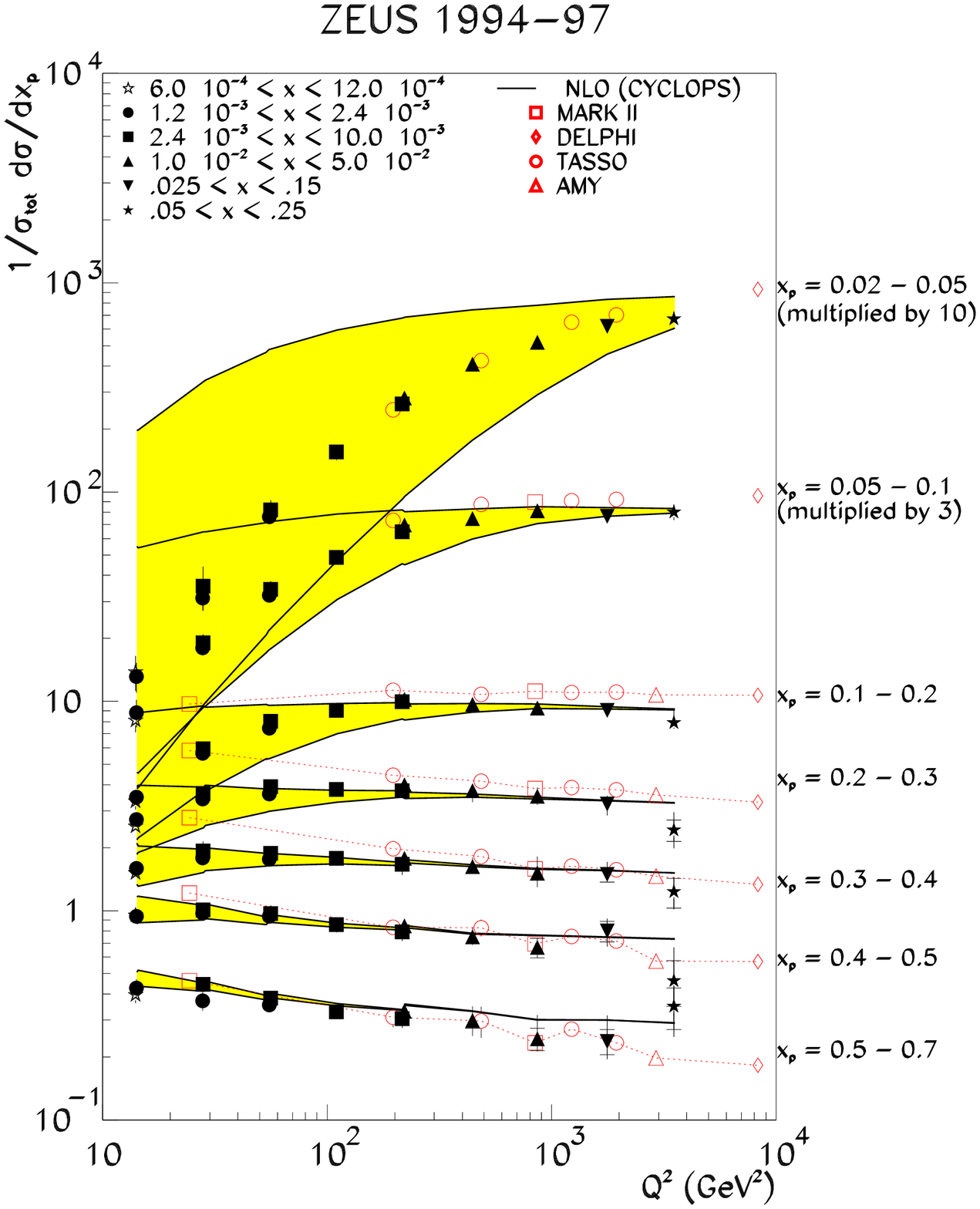,width=70mm}
\caption{Scaling violation in DIS fragmentation.}
\label{fig:Zeus_537_7ee}\end{center}\end{figure}

On the other hand, the data lie well {\em below} the fixed-order
perturbative prediction \cite{Graudenz:1997an} at low $x_p$ and $Q^2$
(fig.~\ref{fig:Zeus_537_7ee}).
Discrepancies could be due to power-suppressed ($1/Q^2$)
corrections, of dynamical and/or kinematical origin.
The bands in fig.~\ref{fig:Zeus_537_7ee}
correspond to an ad-hoc correction factor
$$
\left[1+\left(\frac{m_{\mbox{\scriptsize eff}}}{Qx_p}\right)^2\right]^{-1}
\qquad (0.1 <m_{\mbox{\scriptsize eff}}< 1 \mbox{ GeV}).$$

\item In the {\em target hemisphere} there is also disagreement with
MLLA \cite{Breitweg:1999nt},
possibly due to the T3 contribution ``leaking'' into the region $x_p<1$.
If anything, Monte Carlo models predict too much leakage
(fig.~\ref{fig:Zeus_537_13}).
Little $Q^2$ dependence is evident.
\begin{figure}\begin{center}\epsfig{file=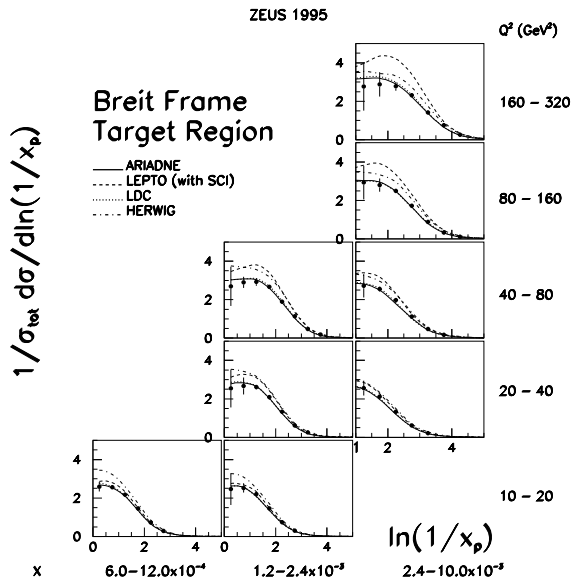,width=8cm}
\caption{Target fragmentation compared with models.}
\label{fig:Zeus_537_13}\end{center}\end{figure}
\end{itemize}

\section{Heavy quark fragmentation}\label{sec:heavy}
\begin{figure}\begin{center}\epsfig{file=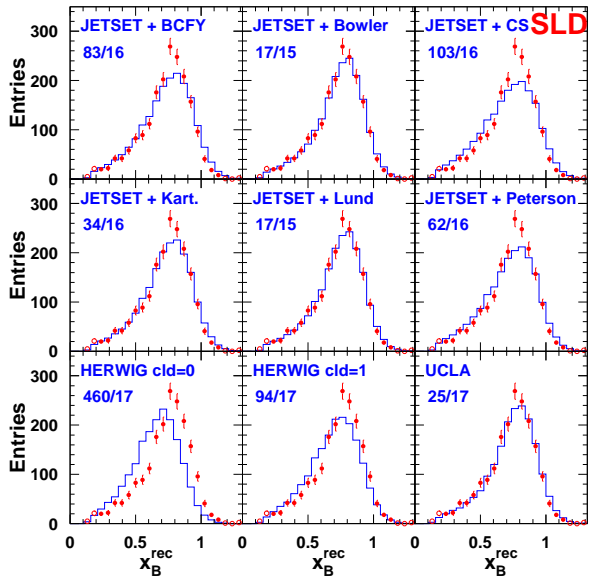,width=80mm}
\caption{SLD data on $b\to$ B fragmentation compared with models.}
\label{fig:SLD_8153_8}\end{center}\end{figure}

New data on $b\to$ B fragmentation from SLD \cite{Abe:1999fi},
using high-precision vertexing, discriminate between parton-shower
plus hadronization models (fig.~\ref{fig:SLD_8153_8}).
Note that the data have not yet been corrected for detector effects.

Including more perturbative QCD leads to a reduction in the amount of
non-perturbative smearing required to fit the data. Non-perturbative
effects are conventionally parametrized by $\eps_b$ in the Peterson
function \cite{Peterson:1983ak}
$$f(z)= \frac 1z\left(1-\frac 1z-\frac{\eps_b}{1-z}\right)^{-2}
\;\;\;\;\;\;\;(z=x_B/x_b)$$

\begin{figure}\begin{center}
\begin{minipage}{50mm}
\epsfig{file=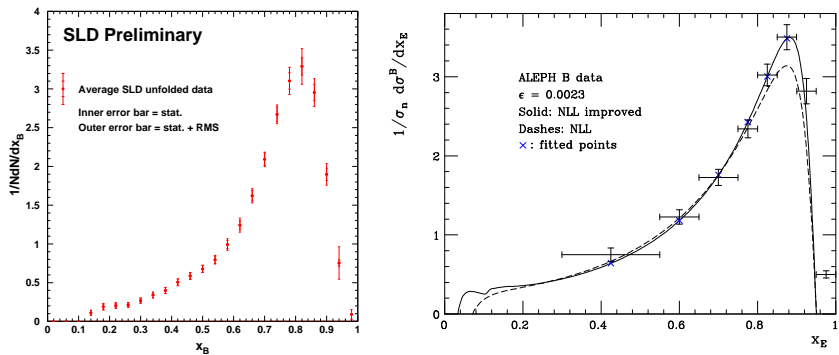,width=5cm}\end{minipage}
\begin{minipage}{50mm}
\epsfig{file=aleph_nllimprov.eps,width=6cm}\end{minipage}
\caption{SLD \cite{Abe:1999fi} and ALEPH \cite{Buskulic:1995gp}
data on $b\to$ B fragmentation, the latter compared
with NLLA QCD \cite{Nason:1999zj}.}\label{fig:aleph_nllimprov}
\end{center}\end{figure}

\noindent
Pure Peterson \cite{Abe:1999fi}: $\eps_b= 0.036$.\\
JETSET ($\simeq$ LLA QCD) + Peterson \cite{Abe:1999fi}: $\eps_b= 0.006$.\\
NLLA QCD + Peterson \cite{Nason:1999zj}: $\eps_b= 0.002$
(fig.~\ref{fig:aleph_nllimprov}).

In the universal low-scale $\as$ model, the perturbative prediction
is extrapolated smoothly to the non-perturbative region, with no Peterson
function at all \cite{Dokshitzer:1996ev}.

\section{Bose-Einstein correlations}\label{sec:BE}
\begin{figure}\begin{center}\epsfig{file=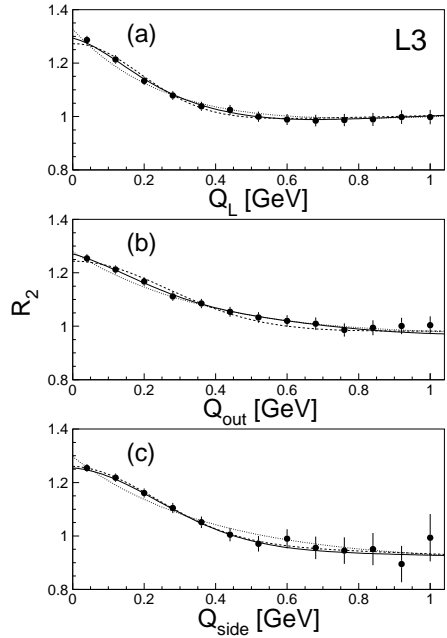,width=6cm}
\caption{Bose-Einstein correlations with repect to axes
along and perpendicular to the thrust axis.}
\label{fig:L3_3_280_3}\end{center}\end{figure}

Studies of  $\pi^\pm\pi^\pm$ correlations which distinguish between
directions along and perpendicular to the thrust axis find
definite evidence for elongation of the
source region along that axis (fig.~\ref{fig:L3_3_280_3} and
table \ref{tab:BErad} \cite{1_221,3_280,3_64}).
This has a good explanation in the Lund string model,
in terms of the change of the space-time area $A$ in
fig.~\ref{fig:string_area} when identical bosons are 
interchanged \cite{Andersson:1998xd}.

\begin{table}\begin{center}{\small
\begin{tabular}{|c|c|c|c|} \hline 
Expt & $R_L$ (fm)&  $R_T$ (fm) & $R_L/R_T$\\ 
\hline
DELPHI & $0.85\pm 0.04$ & $0.53\pm 0.04$ & $1.61\pm 0.10$ \\
L3 & $0.74\pm 0.04$  & $0.56^{+0.03}_{-0.06}$
& $1.23\pm 0.03^{+0.40}_{-0.13}$ \\
OPAL & $0.935\pm 0.029$ & $0.720\pm 0.045$ & $1.30\pm 0.12$ \\
\hline
\end{tabular}}
\caption{Longitudinal and transverse source radii.}\label{tab:BErad}
\end{center}\end{table}

\begin{figure}\begin{center}\epsfig{file=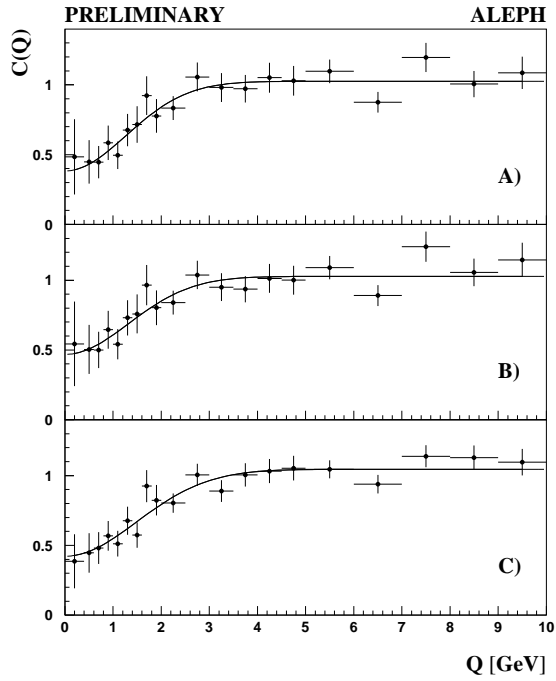,width=8cm}
\caption{Fermi-Dirac correlation for $\Lambda\Lambda$.}
\label{fig:Aleph_1_389_1}\end{center}\end{figure}

ALEPH \cite{1_389} has clear evidence of {Fermi-Dirac} anticorrelation
in $\Lambda\Lambda$ (S=1). Plots A--C in
fig.~\ref{fig:Aleph_1_389_1} correspond to different
comparison (no-correlation) samples.
The source size appears to decrease with increasing particle mass
(table \ref{tab:BEident} \cite{1_389,Decamp:1992md,Abreu:1996hu}).
However, some of this effect is kinematic \cite{Smith:1999bm}.
\begin{table}\begin{center}{\small
\begin{tabular}{|c|c|} \hline 
Particles &  $R_{\mbox{\scriptsize source}}$ (fm)\\ 
\hline
$\pi\pi$ & $0.65\pm 0.04\pm 0.16$ \\
KK & $0.48\pm 0.04\pm 0.07$ \\ 
$\Lambda\Lambda$ & $0.11\pm 0.02\pm 0.01$ \\
\hline
\end{tabular}}
\caption{Comparison of source radii.}\label{tab:BEident}
\end{center}\end{table}

\section{WW fragmentation}\label{sec:WW}
\begin{figure}\begin{center}\epsfig{file=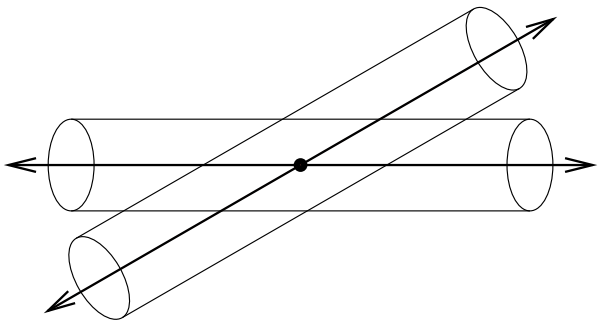,width=8cm}
\caption{Hadronizations volumes in WW decay.}
\label{fig:WWhad}\end{center}\end{figure}

\begin{figure}\begin{center}\epsfig{file=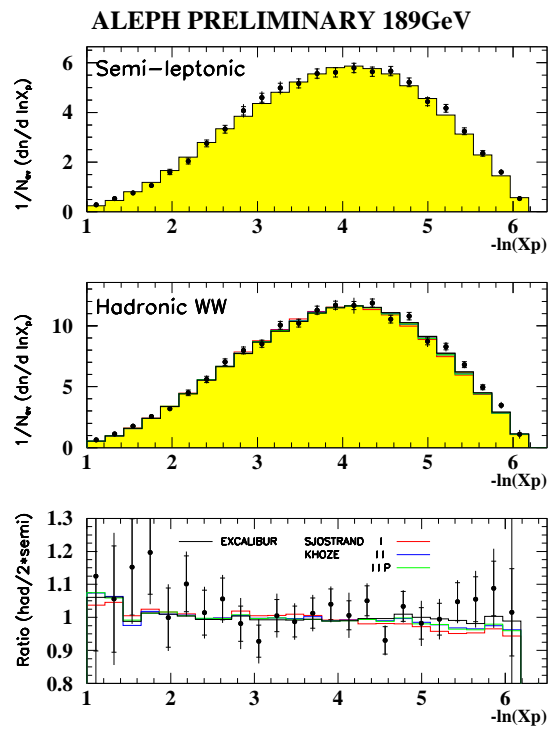,width=8cm}
\caption{Distribution of momentum fraction in WW decay.}
\label{fig:Aleph_1_387_1}\end{center}\end{figure}

\begin{figure}\begin{center}
\begin{minipage}{50mm}
\epsfig{file=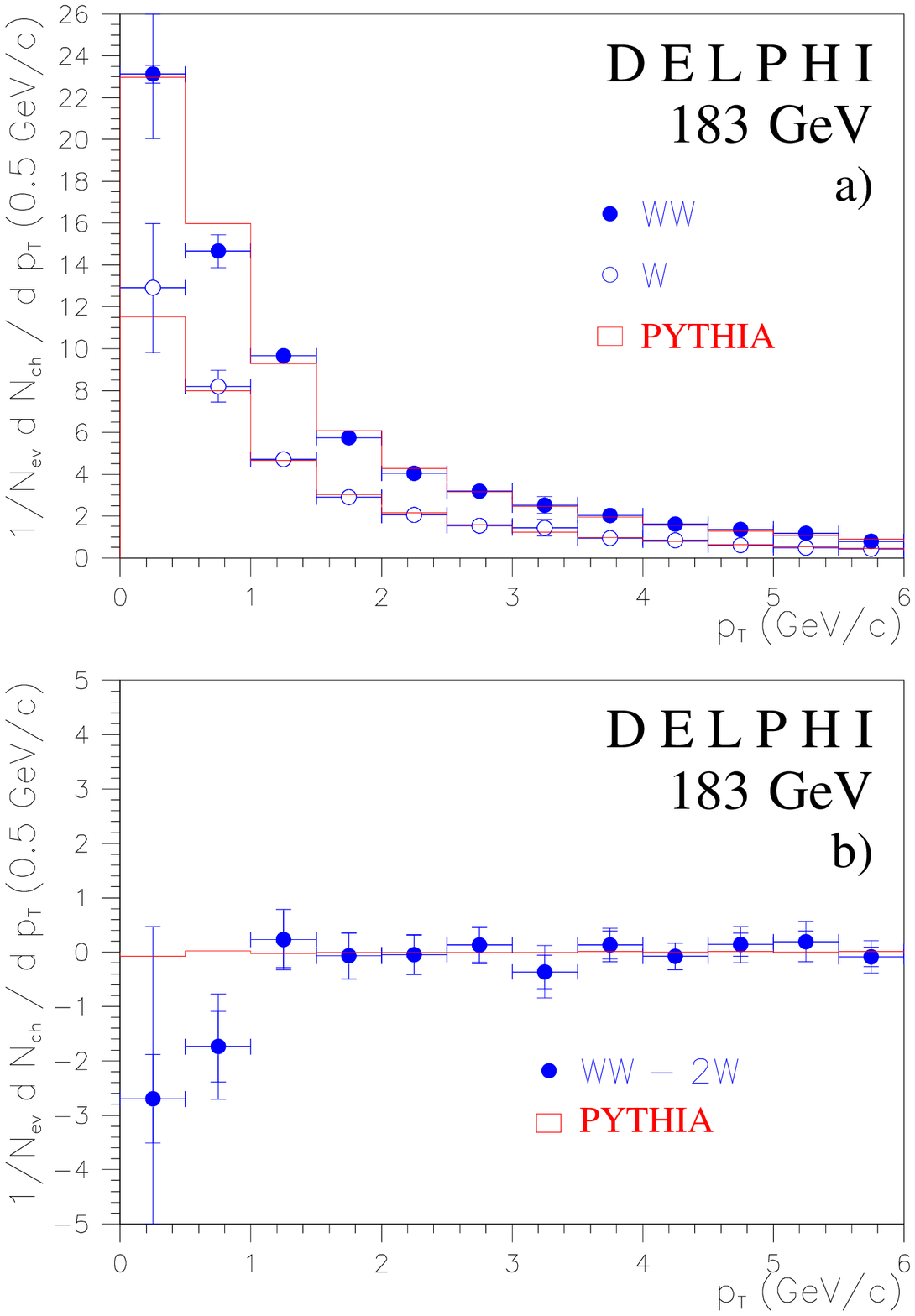,width=5cm}\end{minipage}
\begin{minipage}{50mm}
\epsfig{file=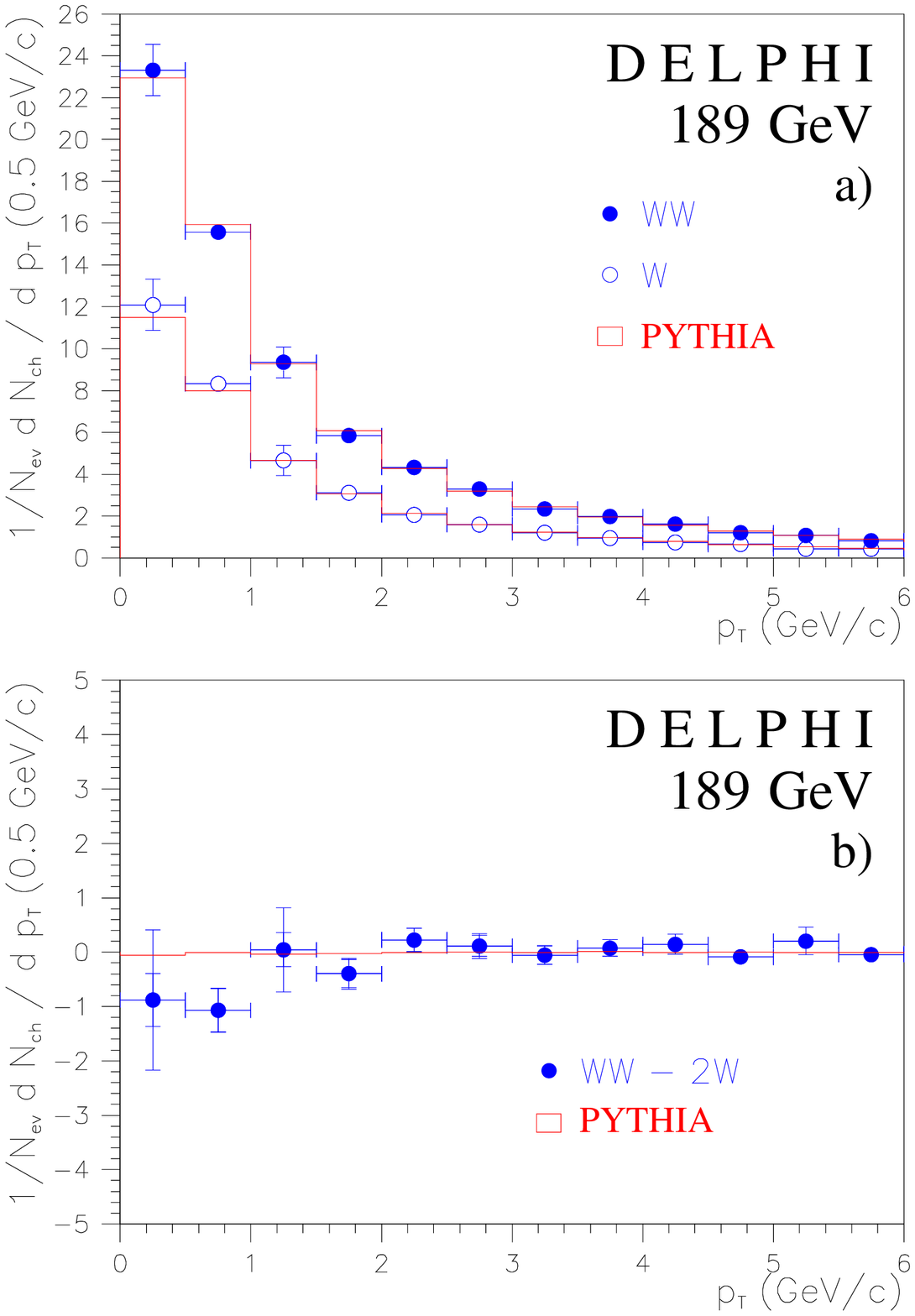,width=5cm}\end{minipage}
\caption{Distribution of transverse momentum in WW decay.}
\label{fig:Delphi_1_229_8}\end{center}\end{figure}

In $\ee\to$ WW, we would expect correlations between W hadronic decays
due to overlap of hadronization volumes.  This occurs mainly
in the central region, and is orientation-dependent
(fig.~\ref{fig:WWhad}).
These {\em reconnection effects} have been searched for
in single-particle distributions.  One would expect
discrepancies between the distributions in semi-leptonic
and fully hadronic decays, especially at low momenta.
There is no firm evidence yet for such effects in the $x_p$ distribution
(fig.~\ref{fig:Aleph_1_387_1} \cite{1_387}). 
However, DELPHI \cite{1_229} report a possible small ($\sim 2\sigma$?)
effect in the distribution of $p_T$ relative to the thrust axis
(fig.~\ref{fig:Delphi_1_229_8}).

\begin{figure}\begin{center}\epsfig{file=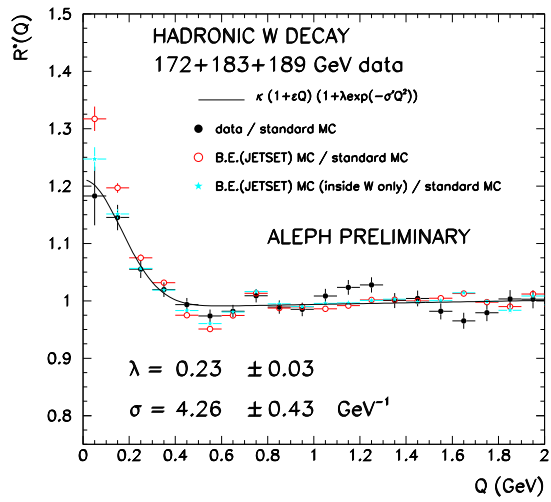,width=8cm}
\caption{Bose-Einstein correlations in WW decay.}
\label{fig:Aleph_1_388_5}\end{center}\end{figure}

{\em Bose-Einstein correlations} between hadrons from different
W's are also being looked for. They would lead to an increase in
the correlation function for WW relative to that for a single W.
There is no sign of any increase at present
(fig.~\ref{fig:Aleph_1_388_5} \cite{1_388}).

\section{Summary}\label{sec:conc}
\begin{itemize}
\item Detailed fragmentation studies need more theoretical input in the
form of coefficient functions that take account of selection procedures,
especially for gluon jets in $\ee$ final states.

\item Hadronization studies suggest that particle masses, rather than quantum
numbers, are the dominant factor in suppressing heavy particle production.
Baryon production is not yet well described by any model.

\item Quark and gluon jets have the expected differences and these can be used
to measure the ratio of colour factors $C_A/C_F$. There is no strong evidence
yet for different particle content in gluon jets.

\item Fragmentation in DIS shows disagreements with perturbative
predictions. It is not yet clear whether these are due to higher-order or
non-perturbative effects.

\item New precise $b$ quark fragmentation data test models and suggest that
perturbative effects dominate.

\item Bose-Einstein (Fermi-Dirac) correlations show elongation of the source
along the jet axis and source shrinkage with increasing mass. 

\item WW fragmentation still shows no firm evidence for correlation between
the decay products of the two W's.
\end{itemize}


\def\Discussion{
\setlength{\parskip}{0.3cm}\setlength{\parindent}{0.0cm}
     \bigskip\bigskip      {\Large {\bf Discussion}} \bigskip}
\def\speaker#1{{\bf #1:}\ }

\Discussion

\speaker{Charles Buchanan (UCLA)}
To elaborate on the UCLA approach to hadronization:  We find that the idea
of a spacetime area law 
(as suggested by strong QCD) works very well as an
organizing principle in the stage of the soft strong-coupled hadronization 
part of the 
process---that is, it gracefully predicts ``easy'' data such as 
light-quark meson production
rates and distributions in $e^+e^-$ with few parameters and 
forms an attractive basis
for studying the relation with the perturbative stage and more 
complicated phenomena such
as baryon formation and $p_T$ effects.  
To study these latter in detail, we have joined
BaBar where we will use the $10^8$ high quality events to be 
collected in the next 2--3 years to study
for baryon-meson-antibaryon 3-body corrections, $p_T$ correlations, etc.
  We invite
physicists interested in the area to contact us 
 (at buchanan@physics.ucla.edu).

\speaker{George Hou (National Taiwan University)}
Regarding particle content of gluon jets, is there any result on the 
$\eta^\prime$ content?  This particle is more naturally associated with
gluons than $\eta$ or $\phi$.

\speaker{Webber}
 As far as I know, there are no results available yet on the $\eta^\prime$
content of gluon jets.

\speaker{Michael Peskin (SLAC)}
You have shown that, when quark and gluon jets are carefully selected, 
their average properties are clearly distinguished.  Of course, 
what one really wants is a variable which
allows one to separate quark and gluon jets (if only statistically) in the 
Tevatron or LHC environment.  What is
the best choice for this purpose?

\speaker{Webber}
 This is difficult to do 
because, although the average properties are different, the
fluctuations are large.

\speaker{Tom Ferbel (University of Rochester)}
Actually, D0 has used the differences between quark and gluon
jets very effectively, in a statistical manner (with neural networks)
 to improve the signal to background
in the analysis of $t\bar t$ production in the all-jets channel.


\end{document}